\documentclass{aa}
\usepackage{graphicx}
\usepackage{afterpage}
\def\tff{t_{\rm ff}}
\def\izw{I\,Zw\,18}
\def\sbs{SBS\,0335$-$052}
\newcommand{\hii}{H\,{\sc ii}}
\begin{document}
\title{Time evolution of the radio continuum of young
starbursts: The importance of synchrotron emission}
\author{H. Hirashita \inst{1}
         \and
L. K. Hunt \inst{2}
}
%
%
\institute{Center for Computational Sciences, University of Tsukuba,
                Tsukuba, Ibaraki 305-8577, Japan\\
\email{hirasita@ccs.tsukuba.ac.jp}
        \and
INAF - Istituto di Radioastronomia-Sezione Firenze, Largo E. Fermi, 5,
50125 Firenze, Italy\\
\email{hunt@arcetri.astro.it}
}
\date{30 August 2006}
\abstract{
We investigate the radio spectral energy distributions
(SEDs) of young star-forming galaxies and how
they evolve
with time. The duration and luminosity of the nonthermal
radio emission from supernova remnants (SNRs) are
constrained by 
using the observational radio SEDs of \sbs\ and \izw, which 
are the two lowest-metallicity blue compact dwarf galaxies
in the nearby universe. 
The typical radio ``fluence'' for SNRs in \sbs, that is the
radio energy emitted
per SNR over its radiative lifetime, is estimated
to be
$\sim 6$--$22\times 10^{22}~{\rm W~Hz^{-1}~yr}$ at 5 GHz.
On the other hand, the radio fluence in \izw\ is
$\sim 1$--$3\times 10^{22}~{\rm W~Hz^{-1}~yr}$ at 5 GHz.
We discuss the origin of this variation and propose
scaling relations between synchrotron luminosity and gas
density. We have also predicted the time dependence of the radio
spectral index and of the spectrum itself, for both
the ``active'' (\sbs) and ``passive'' (\izw) cases.
These models enable us to roughly age date and classify
radio spectra of star-forming galaxies into active/passive
classes. Implications for high-$z$ galaxy evolution are
also discussed.
 }
\titlerunning{Radio continuum of young starbursts}
\authorrunning{H. Hirashita and L. K. Hunt}
\maketitle
%


\section{Introduction}\label{sec:intro}

Radio emission from galaxies is known to be connected with
star formation activity. Two radiative processes are
responsible: thermal free-free radiation
from ionized gas in \hii\ regions
and synchrotron radiation from relativistic electrons
spiraling in magnetic fields (Condon \cite{condon92}).
The former originates from ionized gas around massive stars,
while the latter comes from supernova remnants (SNRs) whose
progenitors are massive stars. Since massive stars have short
lifetimes, both thermal and nonthermal emission should
trace the current star formation rate (SFR).

Nevertheless, the physical basis of the radio emission is
not fully established. 
The observed radio flux in luminous evolved starbursts
and normal spiral galaxies is dominated by nonthermal emission,
but not from the short timescale radiation of discrete SNRs; 
more than $90$\% of it is
probably due to the diffusion of cosmic ray electrons
in the galaxy disk (Condon \cite{condon92}) over timescales of
$10^7$--$10^8$\,yr (Helou \& Bicay \cite{helou93}).
Indeed, if the theoretically estimated adiabatic
timescale (e.g., Woltjer \cite{woltjer72})
and the observationally obtained $\Sigma$--$D$
(surface brightness vs.\ diameter) relation for discrete SNRs 
(e.g., Clark \& Caswell \cite{clark76}) are used to derive 
the nonthermal emission in galaxies 
(e.g., Biermann \cite{biermann76}; Ulvestad \cite{ulvestad82}),
the emission is underpredicted by a factor of 10 or so.

In very young galaxies with ages $\la 10$\,Myr, however, the 
cosmic-ray diffusion mechanisms have not yet had a chance to
dominate the radio emission. Nonthermal synchrotron radiation
from discrete SNRs would be expected
to dominate over the diffuse component.
Moreover, in galaxies with starbursts younger than $\sim 3$\,Myr,
there should be very little synchrotron emission since the SNRs have not
yet exploded. Indeed, some
galaxies with a deficit of nonthermal emission are observed
(Roussel et al.\ \cite{roussel03}) and appear to be due to starbursts
observed within a few Myrs of their onset, after a long period
($\ga$ 100\,Myr) of quiescence.

In this paper, we investigate theoretically the time evolution of the
radio spectral energy distribution (SED) in young
star-forming galaxies.
In principle, this is a straightforward process because we know that
the radio SED can be determined by the star formation history, for
which we have already developed the formalism
(Hirashita et al.\ \cite{hhf02};
Hirashita \& Hunt \cite{hirashita04}).
In practice, it is a difficult exercise because of the unknown
nature of the time evolution of the nonthermal radio component.
Hence, we adopt nearby blue compact dwarf galaxies
(BCDs) as an observational sample to constrain our
theoretical model of radio emission. 
Most BCDs have a young age of {\it current star formation}
(although most of them had previous starburst episodes),
and a low metallicity
(e.g., Izotov \& Thuan \cite{izotov04}).
This implies that they are relatively unevolved chemically
and may provide a reasonable template to investigate star formation
properties of young galaxies.
The BCDs as a class have radio spectra with
different properties than those of evolved luminous star-forming
galaxies (Klein et al.\ \cite{klein91}); flatter radio
spectra of BCDs indicate that they have a lower fraction of
nonthermal emission than more luminous systems.
Hence, they may lack the diffuse emission that characterizes
the radio spectrum of larger disk galaxies, and thus are ideal
targets for constraining our models.

Two classes of star-formation activity in BCDs have recently emerged
observationally, as proposed by
Hunt et al.\ (\cite{hunt-cozumel}) and
Hunt \& Hirashita (2006, in preparation).
They argue that the star-formation modes in the two most
metal-poor galaxies, \object{SBS 0335$-$052} and \object{I Zw 18},
are very different,
in spite of their similar metallicities
($12+\log ({\rm O/H})=7.3$ and $12+\log ({\rm O/H})=7.2$,
respectively;
Skillman \& Kennicutt \cite{skillman93};
Izotov et al.\ \cite{izotovetal99}). The major star-forming
region of \sbs\ is compact and dense (radius
$r_{\rm SF}\la 40$ pc, number density $n\ga 600$ cm$^{-3}$;
Dale et al.\ \cite{dale01}; Izotov \& Thuan \cite{izotov99}).
Moreover, \sbs\ hosts several super star clusters (SSCs),
detectable H$_2$
emission lines in the near-infrared (NIR)
(Vanzi et al.\ \cite{vanzi00}), a large dust extinction
($A_V\sim 16$ mag; Thuan et al.\ \cite{thuan99};
Hunt et al.\ \cite{hunt01};
Plante \& Sauvage \cite{plante02}), and high dust temperature
(Hunt et al.\ \cite{hunt01}; Dale et al.\ \cite{dale01};
Takeuchi et al.\ \cite{takeuchi03}). On the contrary, the
star-forming regions in \izw\ are diffuse
($r_{\rm SF}\ga 100$ pc, $n\la 100$ cm$^{-3}$), and contain
no SSCs. NIR H$_2$ emission has not been detected (Hunt et al.,
private communication), and the dust extinction 
is moderate ($A_V\sim 0.2$ mag; Cannon et al.\ \cite{cannon02}). We
consider a region with such properties as ``passive'' following
Hunt et al.\ (\cite{hunt-cozumel}), while a \sbs-like
star-forming property is considered ``active''.
The similar metallicities of \sbs\ (active) and
\izw\ (passive) imply that
the chemical abundance is not a primary factor in determining
the star-forming properties. We argue that the compactness of
star-forming regions, which affect gas density,
gas dynamics, and so on, is important in the dichotomy
of active and passive modes.

In this paper, we extend this active-passive
classification into the radio regime by focusing on \sbs\ and
\izw\ as  
representatives of metal-poor BCDs, and also of
two star-forming modes in BCDs: ``active'' and ``passive''.
This paper is organized as follows. First, in
Sect.\ \ref{sec:model} we explain the model that describes
the evolution of radio SEDs. Some basic results are given
in Sect.\ \ref{sec:constraint}, where observational data
of \sbs\ and \izw\ are compared with the results for
active and passive modes.
The constraints obtained here are compared with
theoretical descriptions of the physical processes of
synchrotron radiation in Sect.\ \ref{sec:mag}.
In Sect.\ \ref{sec:time} we generalize the results of our
models with predictions for the time evolution of radio SEDs 
distinguishing between active and passive modes,
and discuss implications for high-redshift star formation.
Finally we give our conclusions in Sect.\ \ref{sec:conclusion}.

\section{Model description}\label{sec:model}

The radio emission from galaxies is interpreted to be
composed of two components (Condon \cite{condon92}):
thermal free-free (bremsstrahlung) emission and
nonthermal synchrotron radiation. The former requires
treatment of ionized regions around star-forming regions
and the latter is related to SNRs.
Our models include these processes as described below.
We approximate the dominant star-forming region in
a BCD to a single zone with spherical symmetry.

\subsection{Thermal component}\label{subsec:thermal}

The free-free radiation in ionized regions is responsible
for the flat thermal component of radio SEDs. The
luminosity of the thermal component is proportional to
the number of ionizing photons (with energy larger
than 13.6 eV) emitted per unit time, $\dot{N}_{\rm ion}$.
In order to calculate $\dot{N}_{\rm ion}$, we need the stellar
ionizing photon luminosity and the stellar lifetime.

We use the fitting formulae of Schaerer (\cite{schaerer02})
for the stellar lifetime $\tau_m$ (denoted as $t_\star$
in Schaerer \cite{schaerer02}) and the number of ionizing
photons emitted per unit time $Q(m)$ (see Table 6 of
Schaerer \cite{schaerer02}). Those quantities are
calculated as functions of stellar
mass at the zero age main sequence, $m$.
We adopt $Z=0$ (zero metallicity) for the stellar
properties because we treat a metal-poor phase.
If we adopt the solar metallicity instead,
$\dot{N}_{\rm ion}$ is $\sim 2$ times smaller
(Schaerer \cite{schaerer02}).
Considering those two extreme metallicities, we consider
that the uncertainty caused by the stellar metallicity
is within a factor of 2.

The evolution of $\dot{N}_{\rm ion}$ as a function of time
$t$ is calculated by
\begin{eqnarray}
\dot{N}_{\rm ion}(t)=\int_0^\infty{\rm d}m\int_0^{\tau_m}
{\rm d}t'\, Q(m)\,\phi (m)\,\psi (t-t')\, ,
\end{eqnarray}
where $\psi (t)$ is the SFR at $t$, and $\phi (m)$ is the
initial mass function (IMF; the definition of the IMF is
the same as that in Tinsley \cite{tinsley80}). In this
paper, we assume a
Salpeter IMF $\phi (m)\propto m^{-2.35}$
(Salpeter \cite{salpeter55}) with stellar mass
range of 0.1--100 $M_\odot$. With this IMF, the
mass fraction of stars more massive than 5 $M_\odot$
is 0.18. The calculated radio luminosities are roughly
scaled with this fraction of massive stars if another IMF
is adopted. The functional form of the SFR $\psi (t)$ is
specified in Sect.\ \ref{subsec:sfr}.

The number of ionizing photons can be related to the
thermal radio luminosity at the frequency $\nu$,
$L_{\rm th}^0(\nu )$ as (Hunt et al.\ \cite{hunt04a},
hereafter H04;
valid for $10000~{\rm K}\la T\la 20000~{\rm K}$)
\begin{eqnarray}
L_{\rm th}^0(\nu ) & = & 1.58\times 10^{19}\,
\frac{n({\rm H}^+)+n({\rm He}^+)}{n({\rm H}^+)}\left(
\frac{\nu}{1\,{\rm GHz}}\right)^{-0.1}\nonumber\\
& & \times\left(
\frac{T}{10^4~{\rm K}}\right)^{-0.31}\left(
\frac{\dot{N}_{\rm ion}}{10^{52}~{\rm s}^{-1}}
\right)~{\rm W~Hz}^{-1}\, ,
\end{eqnarray}
where $n({\rm H}^+)$ and $n({\rm He}^+)$ are the number
densities of ionized hydrogen and ionized helium,
respectively (we assume $n({\rm He}^+)=0.08n({\rm H}^+)$
in this paper; H04), and $T$ is the gas temperature.
For the gas temperature in the ionized region, we assume
a typical value of metal-poor \hii\ regions:
$T\simeq 2\times 10^4$ K
(Skillman \& Kennicutt \cite{skillman93};
Izotov et al.\ \cite{izotov99}). The change of
the temperature from $2\times 10^4$ K to
$1\times 10^4$ K has only a minor effect on
$L_{\rm th}^0$ and causes a factor of 2.5 difference
in the free-free optical depth estimated below
(Eq.\ \ref{eq:tau_ff}).

A part of the radio emission is self-absorbed by the ionized gas.
We estimate the optical depth  of the free-free
radiation, $\tau_{\rm ff}$ as
\begin{eqnarray}
\tau_{\rm ff} \simeq 0.328\left(
\frac{T}{10^4\,{\rm K}}\right)^{-1.35}\left(
\frac{{\rm EM}}{10^6\,{\rm pc\, cm}^{-6}}\right)
\left(\frac{\nu}{\rm GHz}\right)^{-2.1}\label{eq:tau_ff}
\end{eqnarray}
(H04), where the emission measure ${\rm EM}$ is estimated by
assuming a constant electron number density $n_e$:
${\rm EM}=n_e^2\ell$ ($\ell$ is the length of the ionized
region along the line of sight). The length $\ell$ is
estimated by dividing the volume with the surface
of the ionized region.
We thus obtain the emission measure as
\begin{eqnarray}
{\rm EM}=\frac{4}{3}n_{e}^2r_{\rm i}\, ,\label{eq:em}
\end{eqnarray}
 where $r_{\rm i}$ is
the radius of ionized region defined in
Sect.\ \ref{subsec:ri}.
In the ionized region, the following condition is
assumed to be satisfied for the charge neutrality
(H04):
\begin{eqnarray}
n_{\rm e}=n({\rm H}^+)+n({\rm He}^+)\simeq 1.08n_{\rm H}\,
,\label{eq:corr_he}
\end{eqnarray}
where $n_{\rm H}$ is the number of
hydrogen nuclei.
Since the ionized gas is responsible for both emission
and absorption, it is reasonable
to assume that the absorbing medium is intermixed
with the emitting material. In this case, the
monochromatic luminosity of the thermal component
emerging from the ionized region becomes
\begin{eqnarray}
L_{\rm th}(\nu )=
\frac{1-\exp (-\tau_{\rm ff})}{\tau_{\rm ff}}\,
L_{\rm th}^0(\nu )\, .
\end{eqnarray}

\subsection{Nonthermal component}\label{subsec:nonthermal}

Nonthermal radio emission from galaxies originates from
SNRs. Since we consider young star-forming regions,
we only
treat Type II supernovae (SNe II), whose progenitors have
short lifetimes\footnote{Although see Mannucci et al.\
(\cite{mannucci05}) for the idea that Type Ia supernovae are 
important in young systems.}. 
The rate of SNe II as a function of time,
$\gamma (t)$, is given by
\begin{eqnarray}
\gamma (t)=\int_{8~M_\odot}^{\infty}\psi (t-\tau_m)\,
\phi (m)\,{\rm d}m\, .\label{eq:sn2}
\end{eqnarray}

First we use a ``standard'' estimate for the intrinsic
nonthermal radio luminosity at frequency $\nu$,
$L_{\rm nt,s}^0(\nu )$ by relating it to $\gamma(t)$
(e.g., Condon \cite{condon92}):
\begin{eqnarray}
L_{\rm nt,s}^0(\nu ) & = & 1.2\times10^{22}\left(
\frac{E_{\rm SN}}{10^{51}~{\rm erg}}\right)^{-1/17}
\left(\frac{n_{\rm H}}{1~{\rm cm}^{-3}}\right)^{-2/17}
\nonumber\\
& \times & \left(\frac{\gamma (t)}{1~{\rm yr}^{-1}}
\right)\left(\frac{\nu}{408~{\rm MHz}}\right)^{\alpha}~
{\rm W~Hz}^{-1}\, ,\label{eq:nt_condon}
\end{eqnarray}
where $E_{\rm SN}$ is the explosion energy of a SN II
($E_{\rm SN}=10^{51}$ erg is assumed in this paper),
and $\alpha$ is the spectral index. We designate this
model for the nonthermal component 
as the ``s-model'' (``s'' stands for standard).
However, this formula is obtained by assuming the
$\Sigma -D$ (surface brightness vs.\ diameter) relation
derived from the Galactic SNRs
(Clark \& Caswell \cite{clark76};
Ulvestad \cite{ulvestad82}) and the adiabatic
lifetime $\tau_{\rm ad}$ as time during which the
remnant emits in the radio. 
The adiabatic timescale is estimated as
(Woltjer \cite{woltjer72}; Condon \cite{condon92}):
\begin{eqnarray}
\tau_{\rm ad}\simeq 3.4\times 10^4\left(
\frac{E_{\rm SN}}{10^{51}~{\rm erg}}\right)^{4/17}
\left(\frac{n_{\rm H}}{1~{\rm cm}^{-3}}\right)^{-9/17}
\,~{\rm yr}\, .
\label{eq:adiabatic}
\end{eqnarray}
The SN II rate inferred from the nonthermal radio
luminosities by using Eq.\ (\ref{eq:nt_condon})
is generally too high (Condon \cite{condon92}
and references therein; see also
Sect.\ \ref{sec:intro}). In other words, the
radio luminosity given by the dependence on $\gamma$
in Eq.\ (\ref{eq:nt_condon})
is an underestimate. Thus, we need to reexamine the
underlying physical assumptions. 
Condon (\cite{condon92}) suggests that the discrepancy
arises because SNRs produce more than 90\% of the
nonthermal radio emission long after the individual
SNRs have faded out and the electrons have diffused
throughout the galaxy. However, this may not be true
in young starbursts, hence warranting an
investigation of the possibility that individual SNRs have
a radio lifetime longer than the adiabatic timescale
(Ilovaisky \& Lequeux \cite{ilovaisky72}).

In order to constrain the radio energy emitted by a SNR
over its entire lifetime -- the ``fluence'' or
time-luminosity integral -- we treat the duration
($\tau_{\rm nt}$) and the luminosity at a frequency
$\nu_0$ ($l_{\rm nt}$) as parameters to be determined
from observational constraints. Then, the nonthermal
luminosity is written as
\begin{eqnarray}
L_{\rm nt,p}^0(\nu )=l_{\rm nt}\tau_{\rm nt}\gamma
(\nu /\nu_0)^{\alpha}\, .\label{eq:nt_para}
\end{eqnarray}
This model of the nonthermal component is called
``p-model'' (``p'' stands for parameterization).
We take $\nu_0=5~{\rm GHz}$ unless otherwise stated.
We adopt a typical spectral index of the Galactic
SNRs: $\alpha =-0.5$ (Clark \& Caswell \cite{clark76}).
Hunt et al.\ (\cite{hunt05}) mention that
$\alpha =-0.4$ and $-0.5$ fit the radio spectrum of
\izw\ but that the fit is significantly degraded for
$\alpha =-0.8$.

After accounting for the absorption, the nonthermal
radio component emitted by the star-forming region
is estimated as
\begin{eqnarray}
L_{\rm nt}(\nu )=E(\tau_{\rm ff})
L_{{\rm nt},i}^0(\nu )\, ,
\end{eqnarray}
where $E(\tau_{\rm ff})$ is the escaping fraction of
the radio photons, and $i$ is the label indicating
``s'' for the s-model and ``p'' for the p-model.
We consider two cases for $E(\tau_{\rm ff})$ depending
on the geometry of
the absorbing material relative to the radiation
source. One is the ``screen geometry'', where the
absorbing material surrounds the radiation source,
and the other is the ``mixed geometry'', where
the source and the absorber are mixed homogeneously:
\begin{eqnarray}
E(\tau_{\rm ff})=\left\{
\begin{array}{ll}
E_{\rm sc}(\tau_{\rm ff})\equiv\exp (-\tau_{\rm ff})
& \mbox{(screen),} \\
E_{\rm mix}(\tau_{\rm ff})\equiv
{\displaystyle
\frac{1-\exp (-\tau_{\rm ff})}{\tau_{\rm ff}}
}
& \mbox{(mixed).}
\end{array}
\right.
\label{eq:screen}
\end{eqnarray}

\subsection{Star formation rate}\label{subsec:sfr}

Stars form as a result of the gravitational collapse of a gas
cloud. Therefore, it is physically reasonable to relate the
SFR with the free-fall (or dynamical) timescale of
gas (Elmegreen \cite{elmegreen00}). 
We consider a star-forming region with an initial number
density of neutral hydrogen
$n_{\rm H0}$. The free-fall time, $\tff $, is estimated as
\begin{eqnarray}
\tff =\sqrt{\frac{3\pi}{32G\rho}\,}\simeq
\frac{4.35\times 10^7}{\sqrt{\,\mathstrut n_{\rm H0}\,}}
~{\rm yr}\, ,\label{eq:freefall}
\end{eqnarray}
where $G$ is the gravitational constant, and $\rho$ is the
mass density of the gas. In Eq.~(\ref{eq:freefall}), we have
used $\rho =\mu m_{\rm H}n_{\rm H0}$, where $m_{\rm H}$ is the
mass of a hydrogen atom, and the factor $\mu$ is for the
correction for helium. We assume that $\mu =1.4$ in this
paper.

The SFR, $\psi (t)$, basically scales with the gas mass
divided by the free-fall time:
\begin{eqnarray}
\psi (t)=\frac{\epsilon_{\rm SF}M_{\rm gas}}{\tff}f(t)\, ,
\label{eq:sfr}
\end{eqnarray}
where $\epsilon_{\rm SF}$ is the star formation efficiency
defined as the conversion efficiency of a gas into stars
over a free-fall time, $M_{\rm gas}$ is the total gas mass
available in the star-forming region, and the nondimensional
function $f(t)$ specifies the
functional form of the SFR. Numerically, the SFR is
estimated as
\begin{eqnarray}
\psi(t) & \simeq & 0.230\left(
\frac{\epsilon_{\rm SF}}{0.1}\right)
\left(\frac{M_{\rm gas}}{10^7~M_\odot}\right)\left(
\frac{n_{\rm H0}}{100~{\rm cm}^{-3}}
\right)^{1/2}\nonumber \\
& \times & f (t)~M_\odot~{\rm yr}^{-1}\, .
\label{eq:SFR_num}
\end{eqnarray}
We set the zero point of time $t$ at the onset of star
formation in the star-forming
region. We define the termination time of star formation
as $t_{\rm i}$ such that $r_{\rm i}(t_{\rm i})=r_{\rm SF}$,
where $r_{\rm SF}$ is the radius of the star-forming region
(i.e., $t_{\rm i}$ is the time when the entire star-forming
region is ionized; see Sect.\ \ref{subsec:ri} for the
definition of $r_{\rm i}$). If the entire region is never
ionized, $t_{\rm i}$ is taken to be infinity. We note that
$\tff <t_{\rm i}$ is satisfied for the conditions adopted
in this paper. Then, we
investigate the following functional form for $f(t)$:
\begin{eqnarray}
f(t)=f_{\rm c}(t)\equiv\left\{
\begin{array}{@{\,}ll@{\,}}
\exp (-\epsilon_{\rm SF}t/\tff ) &
\mbox{if $0\leq t\leq t_{\rm i}$,} \\
0 & \mbox{if $t<0$ or $t>t_{\rm i}$,}\label{eq:continuous}
\end{array}
\right.
\end{eqnarray}
which is called ``continuous SFR''. In order to represent
an extreme starburst, we also adopt another functional
form for the SFR:
\begin{eqnarray}
f(t)=f_{\rm b}(t)\equiv\left\{
\begin{array}{@{\,}ll@{\,}}
s &
\mbox{if $0\leq t\leq\tff$,} \\
0 & \mbox{if $t<0$ or $t>\tff$,}
\label{eq:instantaneous}
\end{array}
\right.
\end{eqnarray}
where the constant $s$ is adjusted to satisfy
\begin{eqnarray}
s & = & \frac{1}{\tff}
\int_0^{t_{\rm i}}\exp (-\epsilon_{\rm SF}t/\tff )\,
{\rm d}t \nonumber \\
& = & \frac{1}{\epsilon_{\rm SF}}[1-
\exp(-\epsilon_{\rm SF}t_{\rm i}/\tff )] \, ,
\label{eq:s}
\end{eqnarray}
where $t_{\rm i}$ is determined by the continuous SFR
model. This assures that the total stellar mass formed
in a star-forming region is the same between the two
models. In the burst model, all the star formation occurs
on a dynamical timescale. For the star formation
efficiency, we assign a value $\epsilon_{\rm SF}=0.1$
(e.g., Inoue et al.\ \cite{inoue00}).

The initial hydrogen number density can be related to
the gas mass in the star-forming region as
\begin{eqnarray}
\frac{4\pi}{3}r_{\rm SF}^3\mu m_{\rm H}n_{\rm H0}=
M_{\rm gas}\, .
\end{eqnarray}
Then, the numerical value of the number density is
estimated as
\begin{eqnarray}
n_{\rm H0}\simeq 69\left(\frac{r_{\rm SF}}{100~{\rm pc}}
\right)^{-3}\left(\frac{M_{\rm gas}}{10^7~M_\odot}
\right)~{\rm cm}^{-3}\, .\label{eq:density}
\end{eqnarray}

\subsection{Radius of ionized region}\label{subsec:ri}

In order to treat the range of evolution variations from deeply 
embedded \hii\ regions to normal \hii\ regions,
it is crucial to include pressure-driven expansion of \hii\ regions. 
To focus on the radio SED, we adopt a simple analytical 
approximation, which divides 
the evolution of an \hii\ region into two
stages: the first stage is the growth of ionizing front
due to the increase of ionizing photons, and the second
is the pressure-driven expansion of ionized gas. The
expansion speed of ionizing front in the first stage
is simply estimated by the increasing rate of the
Str\"{o}mgren radius. The Str\"{o}mgren radius
$r_{\rm S0}$ under the initial density is defined by the
following relation:
\begin{eqnarray}
\frac{4\pi}{3}r_{\rm S0}^3n_{\rm e}n_{\rm H0}
\alpha^{(2)}=\dot{N}_{\rm ion}\, ,
\end{eqnarray}
where $\alpha^{(2)}$ is the recombination coefficient
excluding captures to the ground ($n=1$) level, and
$n_{\rm e}$ is estimated by Eq.\ (\ref{eq:corr_he}) with
$n_{\rm H}=n_{\rm H0}$.
The increase of $r_{\rm S0}$ is
caused by the accumulation of ionizing stars.
Roughly speaking, as long as $\dot{r}_{\rm S0}$
(the increase rate of the Str\"{o}mgren radius) is
larger than the sound speed of ionized gas,
$C_{\rm II}$
(we assume $C_{\rm II}=10$ km s$^{-1}$ in this
paper), the ionizing front propagates before the
system responds hydrodynamically. Therefore,
we neglect the hydrodynamical expansion if
$\dot{r}_{\rm S0}>C_{\rm II}$, and assume that the
radius of ionized region is equal to the
Str\"{o}mgren radius: $r_{\rm i}=r_{\rm S0}$.

Once $\dot{r}_{\rm S}<C_{\rm II}$ is satisfied,
pressure-driven expansion is treated. Since the density
evolves, we define the Str\"{o}mgren radius $r_{\rm S}$
under the current density:
\begin{eqnarray}
\frac{4\pi}{3}r_{\rm S}^3n_{\rm e}n_{\rm H}
\alpha^{(2)}=\dot{N}_{\rm ion}\, .
\end{eqnarray}
In this situation, the growth of the ionizing region
is governed by the pressure of ionized gas and the
luminosity change of the central stars
has minor effects. Therefore, the
following equation derived by assuming
a constant luminosity (Spitzer \cite{spitzer78})
approximately holds:
\begin{eqnarray}
\dot{r_{\rm i}}=C_{\rm II}
\sqrt{\rho_{\rm II}/\rho_{\rm I}}\, ,
\label{eq:expand1}
\end{eqnarray}
where $\rho_{\rm I}$ and $\rho_{\rm II}$ are the
gas densities outside and inside
of the ionized region, respectively. Here,
$\rho_{\rm II}/\rho_{\rm I}$ is approximately
estimated by
\begin{eqnarray}
\frac{\rho_{\rm II}}{\rho_{\rm I}}\simeq
\left(\frac{r_{\rm i0}}{r_{\rm i}}
\right)^{3/2}\, ,\label{eq:expand2}
\end{eqnarray}
where $r_{\rm i0}$ is evaluated by
the Str\"{o}mgren radius under the current
ionizing photon luminosity and the gas
density of $\rho_{\rm I}$.
Since $n_{\rm H}$ is defined in the ionized
region, we relate it with $\rho_{\rm II}$ as
\begin{eqnarray}
\rho_{\rm II}=\mu m_{\rm H}n_{\rm H}\, .
\end{eqnarray}
The density in the neutral region is assumed
to be constant:
\begin{eqnarray}
\rho_{\rm I}=\mu m_{\rm H}n_{\rm H0}\, .
\end{eqnarray}

Equations (\ref{eq:expand1}) and (\ref{eq:expand2})
are numerically integrated to obtain $r_{\rm i}$ as
a function of $t$. For the convenience of
interpretation, some approximate analytical solutions
are described in Appendix \ref{app:analytic}.

The dynamical expansion is treated as long as
$r_{\rm i}<r_{\rm S}$. When the SFR
declines significantly, $r_{\rm S}$ begins to
decrease. Thus, $r_{\rm i}>r_{\rm S}$ can be
realized at a certain time. When $r_{\rm i}>r_{\rm S}$,
$r_{\rm i}$ is replaced with $r_{\rm S}$ and
$n_{\rm H}$ is fixed;
that is, we finish treating the dynamical expansion.

\section{Observational constraints}\label{sec:constraint}

Here we use observations to constrain
some of the physical parameters in
our models. In particular, the radiative energy --time-luminosity
integral-- of
the nonthermal synchrotron component is the most
important parameter to be obtained in this paper.
Two representative metal-poor BCDs, \sbs\ and \izw\
(Hirashita \& Hunt \cite{hirashita04}), are used here.
In our previous works (Hirashita et al.\ \cite{hhf02};
Hirashita \& Hunt \cite{hirashita04}), \sbs\ has been
used as a proxy for a genuinely young galaxy in which
the mass of underlying old population is negligible
(Vanzi et al.\ \cite{vanzi00}). 
The same may be true for \izw\ in which the mass fraction
due to an underlying evolved stellar population ($\ga$2\,Gyr)
is not more than $\sim 20$\% (Hunt et al. \cite{hti03}).
We adopt 54.3 Mpc for the distance of \sbs\
(Thuan et al.\ \cite{thuan97})
and 12.6 Mpc for \izw\ (\"{O}stlin \cite{ostlin00}).

We calculate the time dependence of the radio SED, $L(\nu )$, by
summing the thermal and nonthermal components:
\begin{eqnarray}
L(\nu )=L_{\rm th}(\nu )+L_{\rm nt}(\nu )\, .
\end{eqnarray}
Each of the following three items has two selections
as summarized in Table \ref{tab:model}:

  \begin{table}
     \caption[]{Models of Radio SEDs}
        \label{tab:model}
\begin{tabular}{@{}cccc@{}}\hline
Model & $L_{\rm nt}^0$\,$^{\rm a}$ &
$E(\tau_{\rm ff})$\,$^{\rm b}$ & SFR$^{\rm c}$ \\
\hline
A & s & s & c \\
B & s & s & b \\
C & s & m & c \\
D & s & m & b \\
E & p & s & c \\
F & p & s & b \\
G & p & m & c \\
H & p & m & b \\
\hline
\end{tabular}
\begin{list}{}{}
\item[$^{\mathrm{a}}$] See Section \ref{subsec:nonthermal}.
  The standard and parameterized models are denoted as ``s''
  and ``p'', respectively (Eqs.\ \ref{eq:nt_condon} and
  \ref{eq:nt_para}, respectively).
\item[$^{\mathrm{b}}$] See Section \ref{subsec:nonthermal}.
  The screen and mixed models are denoted as ``s''
  and ``m'', respectively (Eq.\ \ref{eq:screen}).
\item[$^{\mathrm{c}}$] See Section \ref{subsec:sfr}.
  The continuous and burst SFRs are denoted as ``c''
  and ``b'', respectively (Eqs. \ref{eq:continuous} and
  \ref{eq:instantaneous}, respectively).
\end{list}
  \end{table}

\noindent
(i) nonthermal emission of a SNR: standard / parameterized
(Sect.\ \ref{subsec:nonthermal})

\noindent
(ii) optical depth of nonthermal emission: screen / mixed
(Sect.\ \ref{subsec:nonthermal})

\noindent
(iii) SFR: continuous/burst
(Sect.\ \ref{subsec:sfr})

\noindent
In total we have 8 models, which are labeled as
Models A--H as shown in Table \ref{tab:model}.

\subsection{\sbs}\label{subsec:sbs}

We adopt $r_{\rm SF}=20$ pc and
$n_{\rm H0}=7\times 10^3~{\rm cm}^{-3}$ (i.e.,
$M_{\rm gas}=8.1\times 10^6~M_\odot$;
$\tff =5.2\times 10^5$ yr). We selected these
values after surveying the full range of
parameters so that the
observable quantities (SFR, $n_{\rm H}$, and
EM) are consistent with observations at the
age estimated by Vanzi et al.\ (\cite{vanzi00}).
The time evolution of the relevant
quantities is shown below to verify whether those
values indeed reproduce the observations.
The radius of the star-forming region is also consistent with
Takeuchi et al.\ (2005), who
reproduce the observational FIR SED of
\sbs.

\begin{figure*}
\includegraphics[width=8cm]{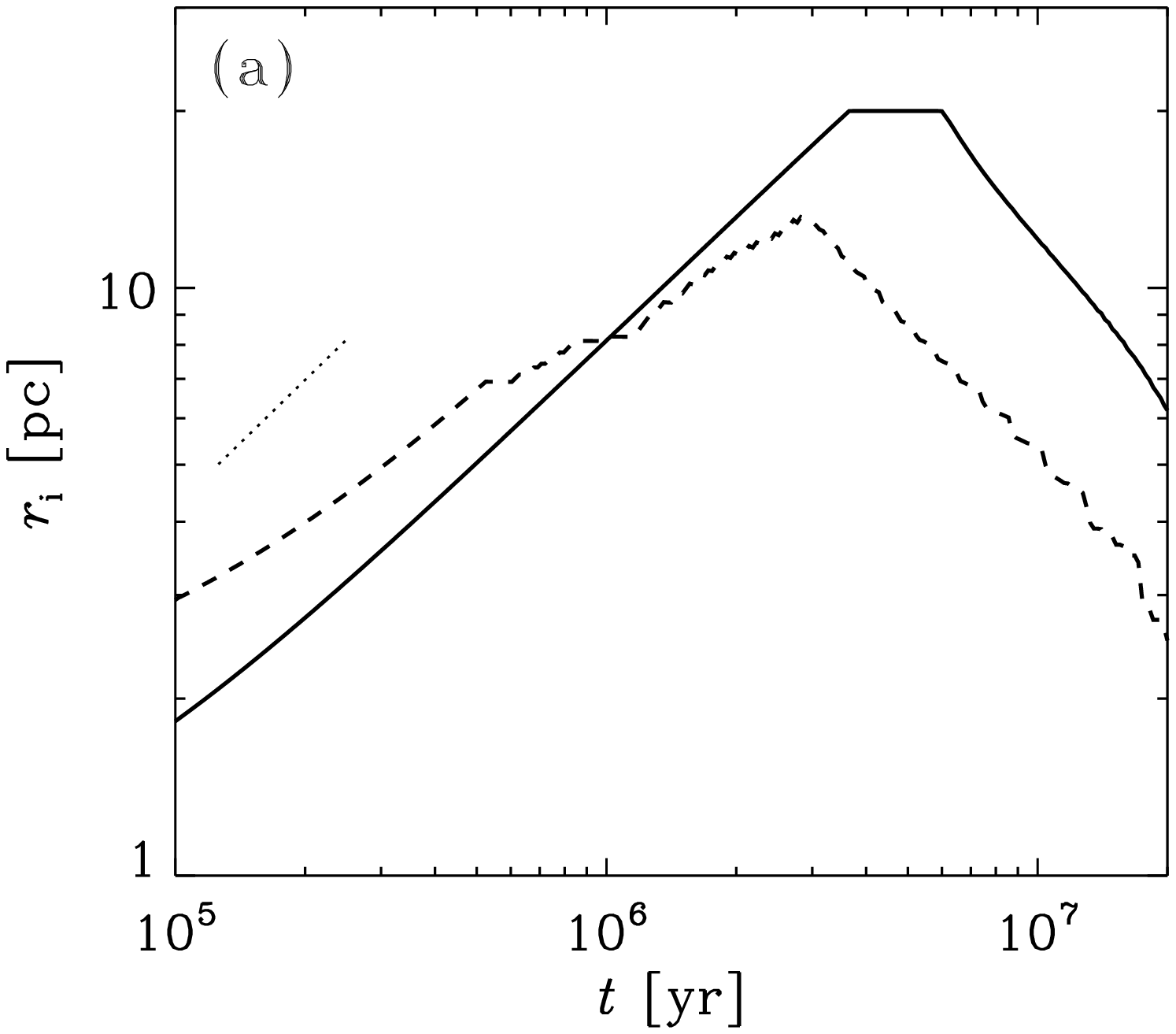}
\includegraphics[width=8cm]{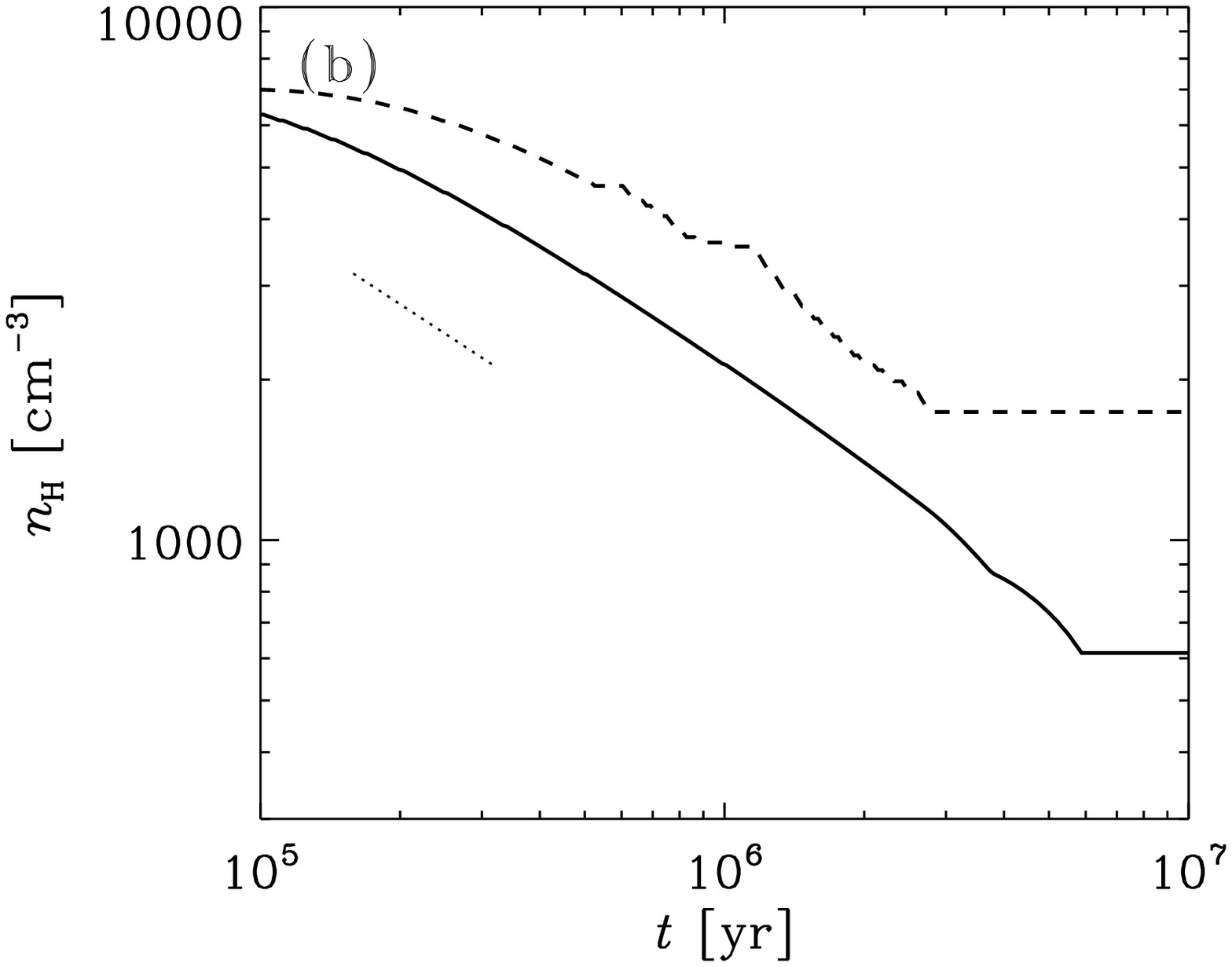}\\
\includegraphics[width=8cm]{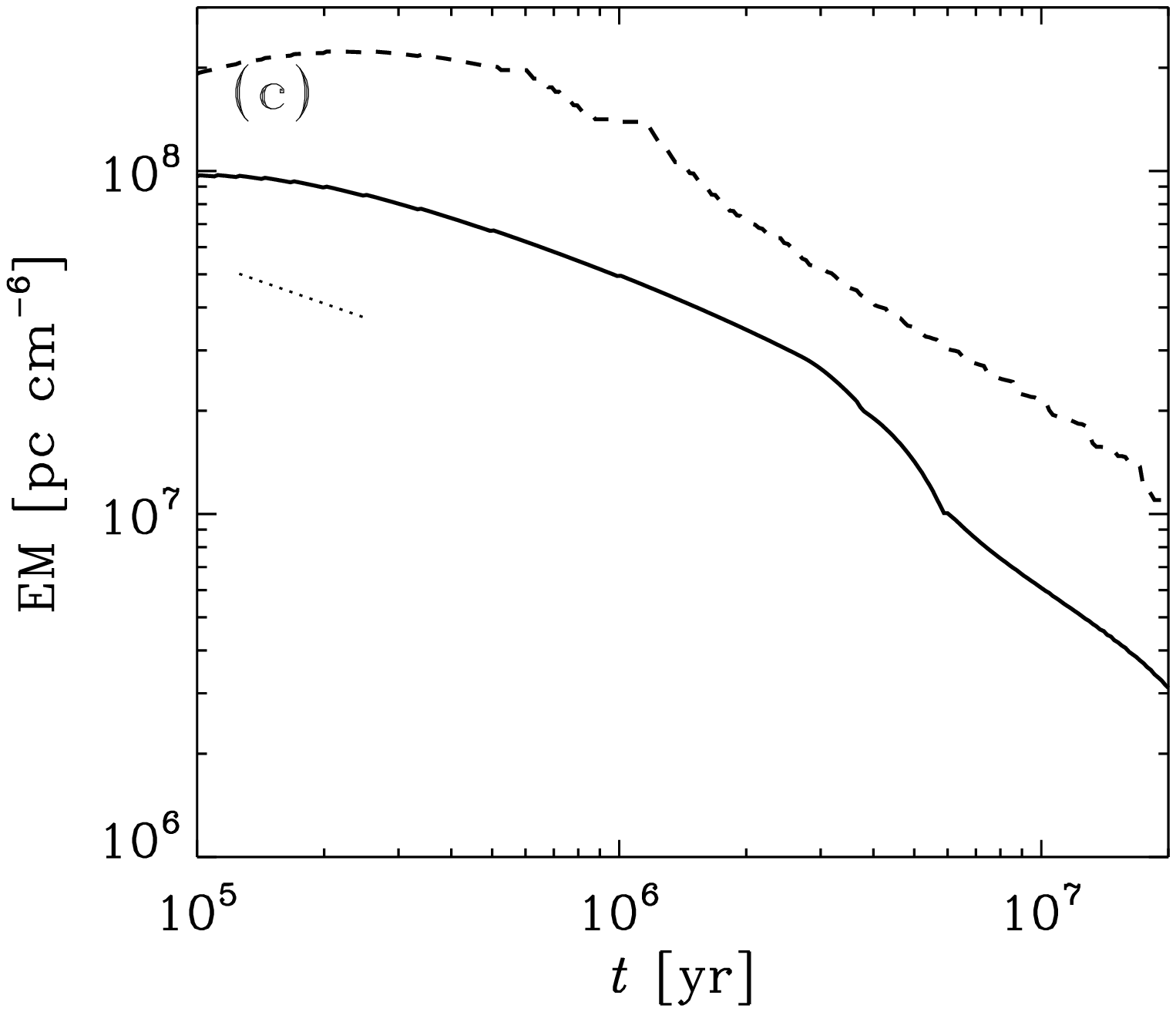}
\includegraphics[width=8cm]{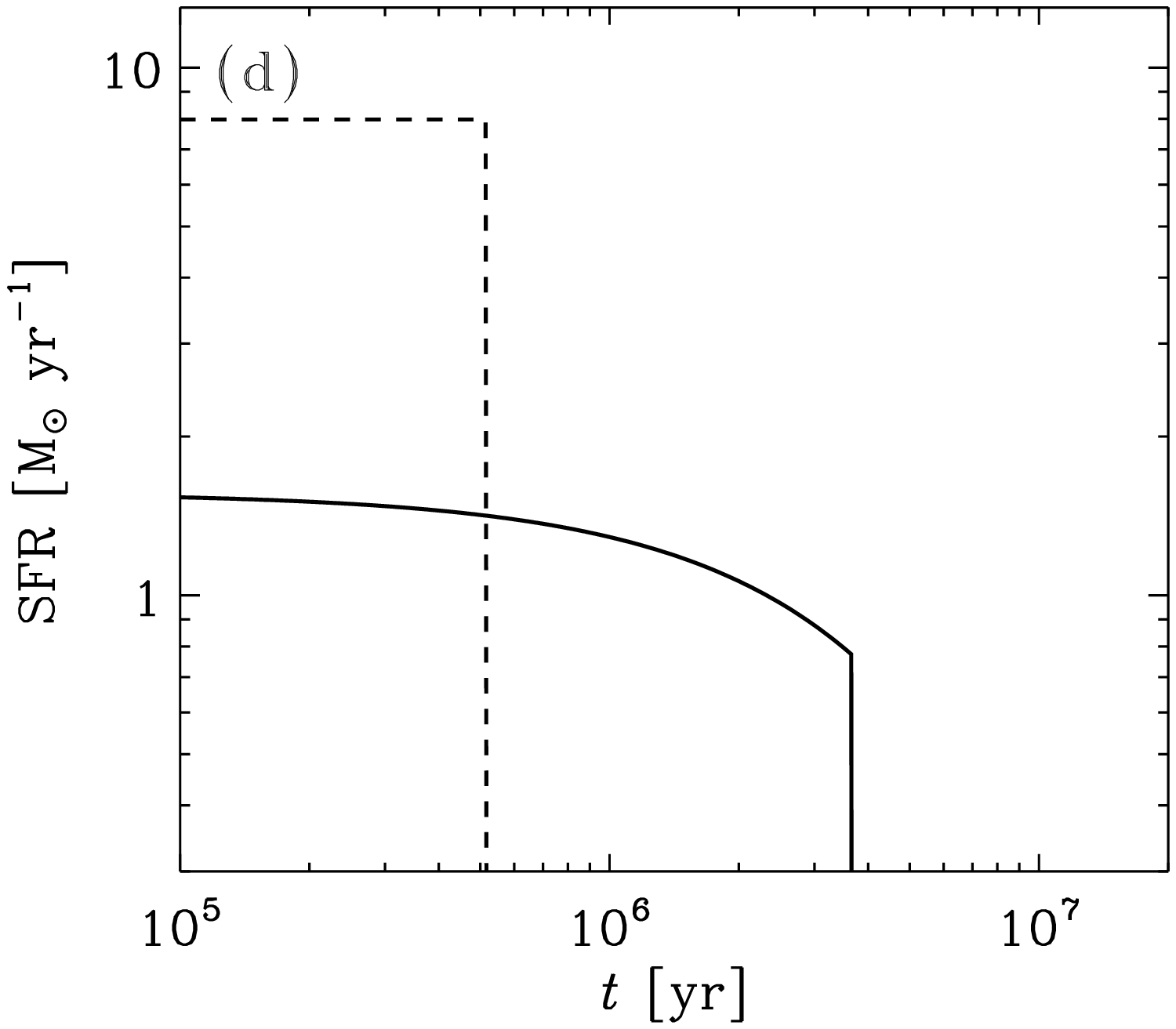}
\caption{Time evolution of basic quantities
concerning radio emission,  
with the solid and dashed lines corresponding to
the continuous and burst SFR, respectively.
The initial conditions are selected to
be the ``active'' mode, consistent with
the observations of \sbs\ at $t\simeq 5$ Myr.
{\bf a)} Radius of ionized region $r_{\rm i}$ as a
function of time. {\bf b)} Number density of
protons in the ionized region $n_{\rm H}$.
{\bf c)} Emission measure.
{\bf d)} Star formation rate.
The dotted lines show the analytical
approximation: $r_{\rm i}\propto t^{5/7}$,
$n_{\rm H}\propto t^{-4/7}$, and
${\rm EM}\propto t^{-3/7}$.}
\label{fig:sbs_basic}
\end{figure*}

First, we evaluate the continuous SFR. The time
evolution of $r_{\rm i}$, $n_{\rm H}$, EM, and SFR 
is shown in Figures \ref{fig:sbs_basic}a, b, c, and d,
respectively. The SFR averaged for 5\,Myr
is $\simeq 0.8~M_\odot~{\rm yr}^{-1}$ with the
total stellar mass formed of $4\times 10^6~M_\odot$.
The SFR of stars more massive than 5 $M_\odot$
(SFR$_{>5~M_\odot}$),
which is less affected by the assumed IMF, is
estimated to be 0.14 $M_\odot$ yr$^{-1}$, which is
consistent with the estimate by H04
(SFR$_{>5~M_\odot}\simeq 0.13$--0.15 $M_\odot$ yr$^{-1}$).
At $t=5$ Myr,
$n_{\rm H}\simeq 7\times 10^2~{\rm cm}^{-3}$, which is
also compatible with the observational results
(Izotov et al.\ \cite{izotov99}; H04).

The radius of the ionized region $r_{\rm i}$ monotonically
increases until $t=3.7$ Myr because of the
pressure-driven expansion and the increase of
$\dot{N}_{\rm ion}$. At $t=3.7$ Myr, the entire
star-forming region is ionized. Until then, the
increasing behavior of $r_{\rm i}$ is approximated by
$r_{\rm i}\propto t^{5/7}$ as derived in
Appendix \ref{app:analytic}. In
Figure \ref{fig:sbs_basic}a, we show the slope of
$r_{\rm i}\propto t^{5/7}$ (dotted line), which
is roughly consistent with the calculated
result (solid line). The initial
evolution of density is also explained by the
analytical relation derived in Appendix as shown
in Figure \ref{fig:sbs_basic}b
($n_{\rm H}\propto t^{-4/7}$). As a result, the
emission measure decreases as
${\rm EM}\propto t^{-3/7}$ as shown in
Figure \ref{fig:sbs_basic}c.

If we adopt the burst SFR ($s=5.1$; Eq.\ \ref{eq:s}),
we obtain the time evolutions as depicted in
Figure \ref{fig:sbs_basic}. As mentioned above,
value of $s$ is
determined to ensure that the total stellar mass
formed is the same between the burst and continuous
modes. Since the star
formation stops on a short timescale, the driving
force of the dynamical expansion decreases more rapidly
than in the case of the continuous SFR. Thus, the 
maximum radius of the ionized region is smaller
and the density is kept higher. This high density
makes the EM higher in the burst SFR, and the EM
derived observationally by H04
(${\rm EM}=3-7\times 10^7$ pc cm$^{-6}$) is more
consistent with the burst SFR than with the
continuous SFR at $t\sim 5$ Myr.
Although the burst scenario is more consistent with
the observations of \sbs, in what follows we examine the 
two cases as extremes for the SFR.

\begin{figure*}[!h]
\centering
\includegraphics[width=8cm]{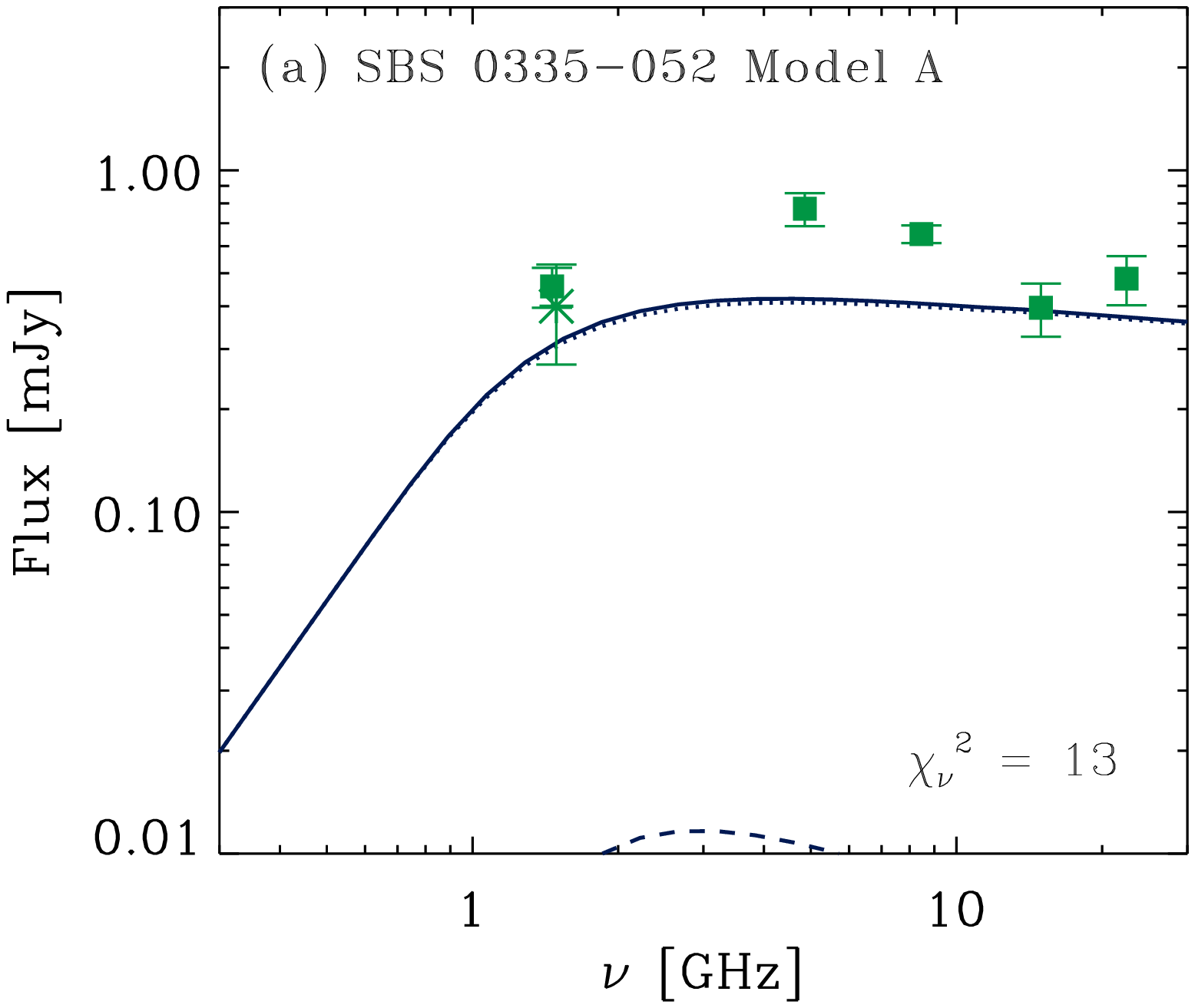}
\includegraphics[width=8cm]{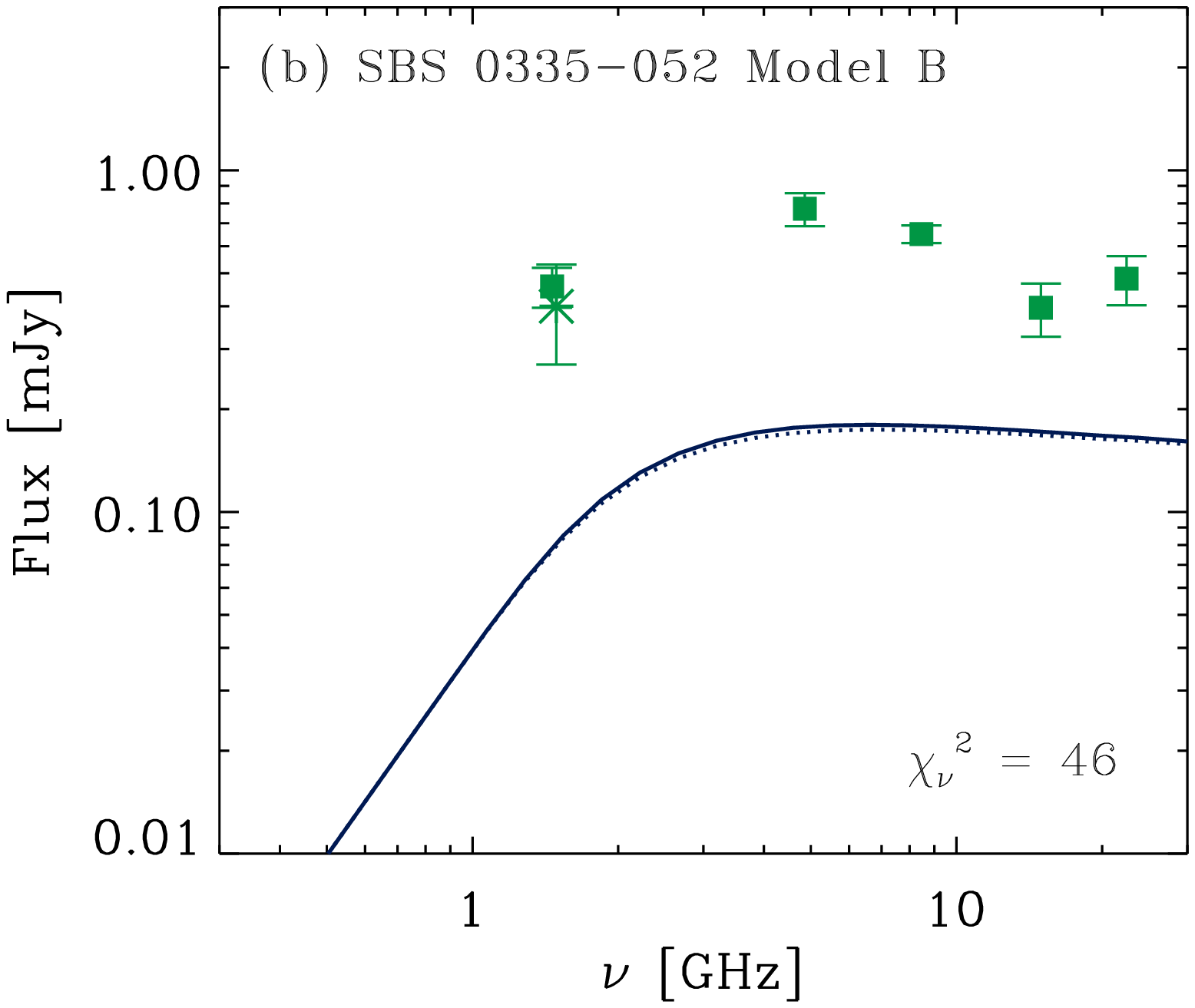}\\
\includegraphics[width=8cm]{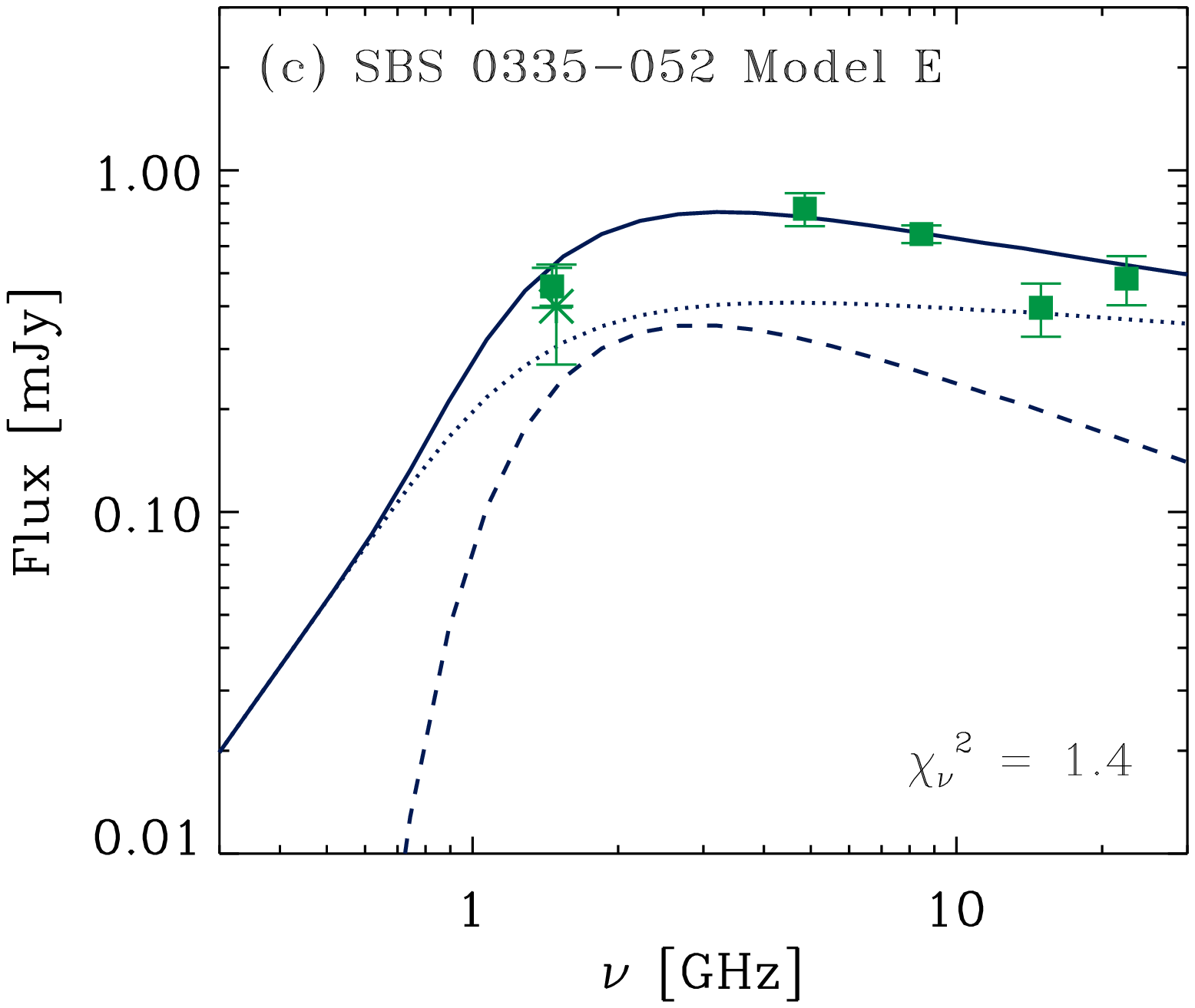}
\includegraphics[width=8cm]{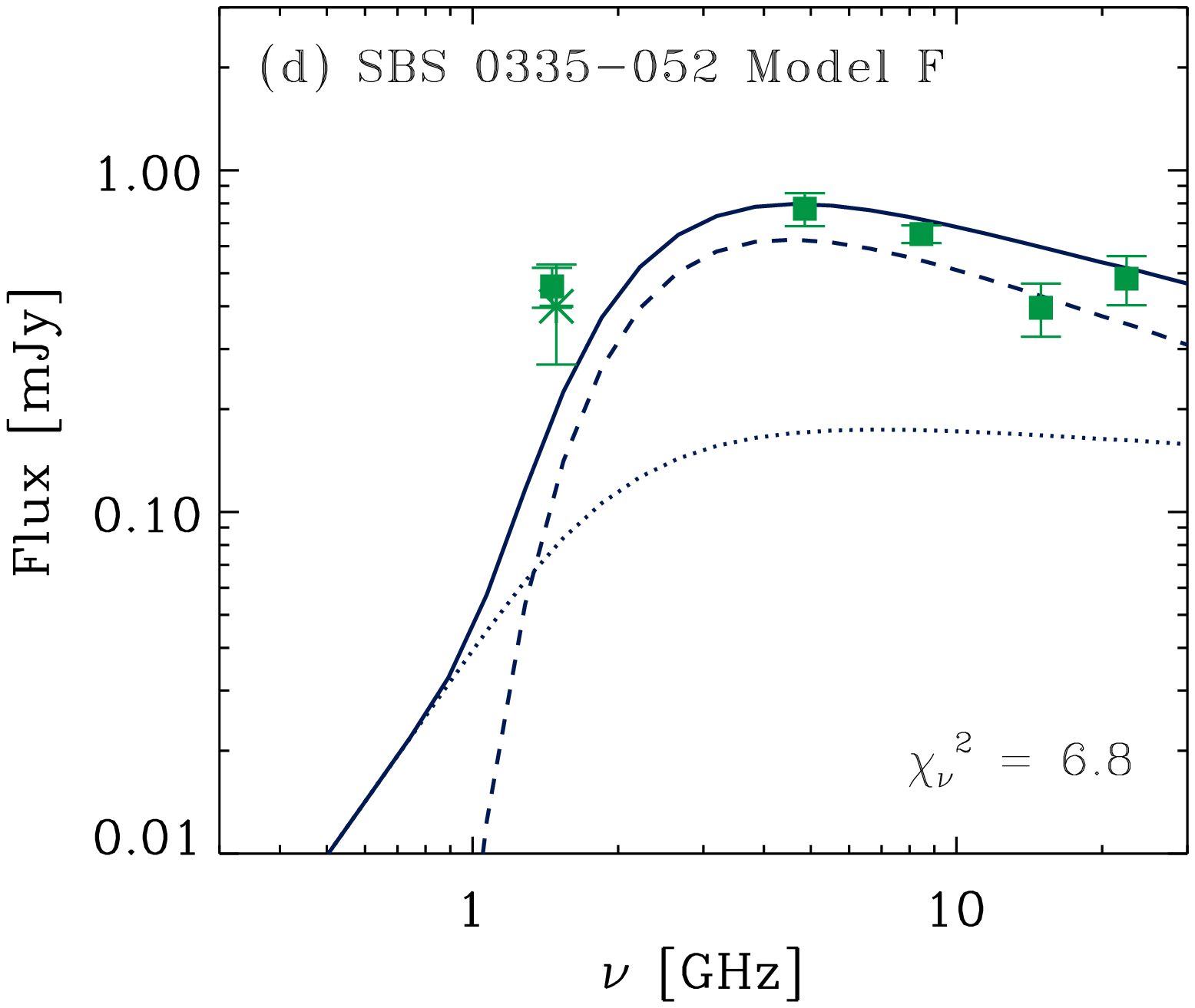}\\
\includegraphics[width=8cm]{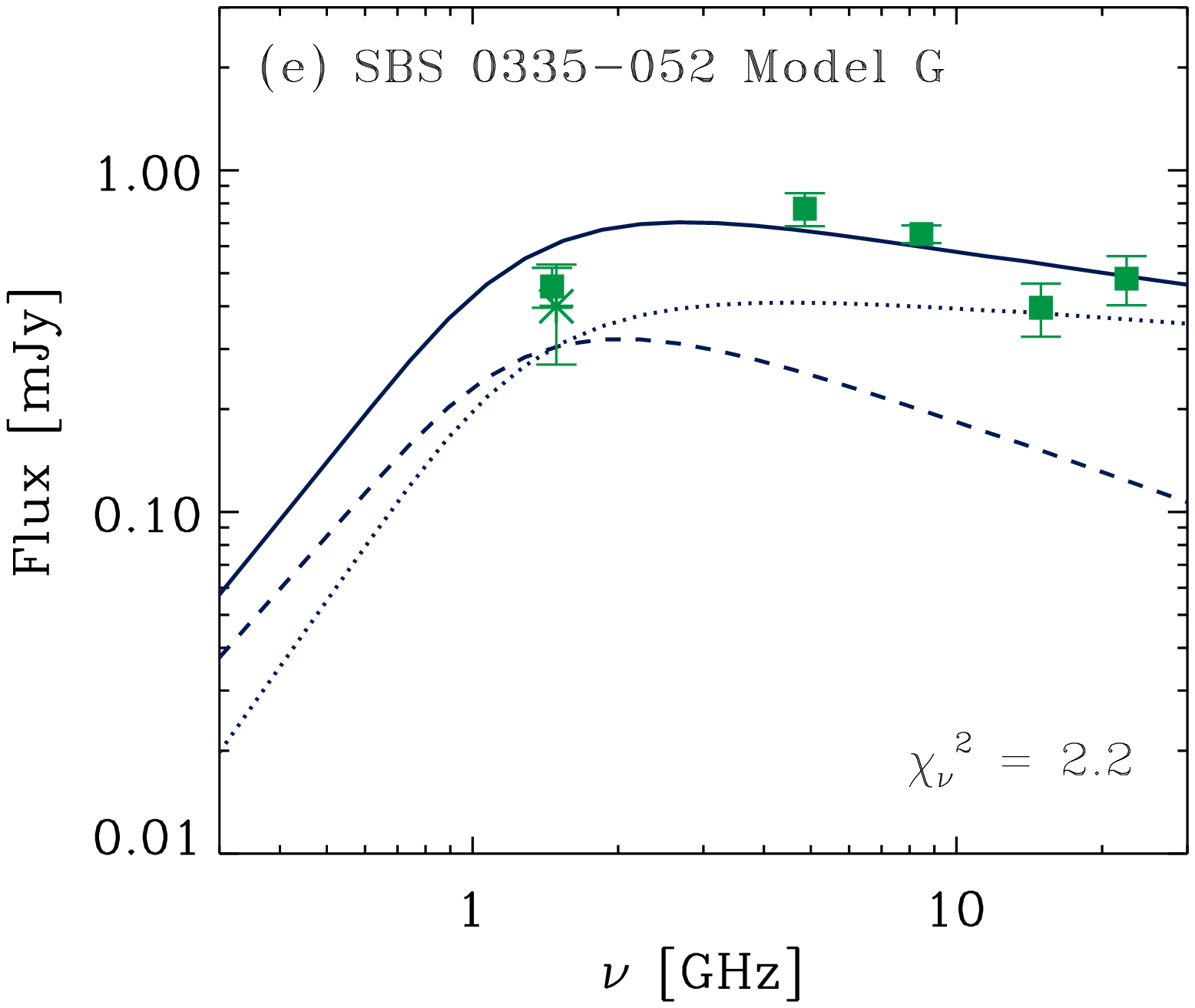}
\includegraphics[width=8cm]{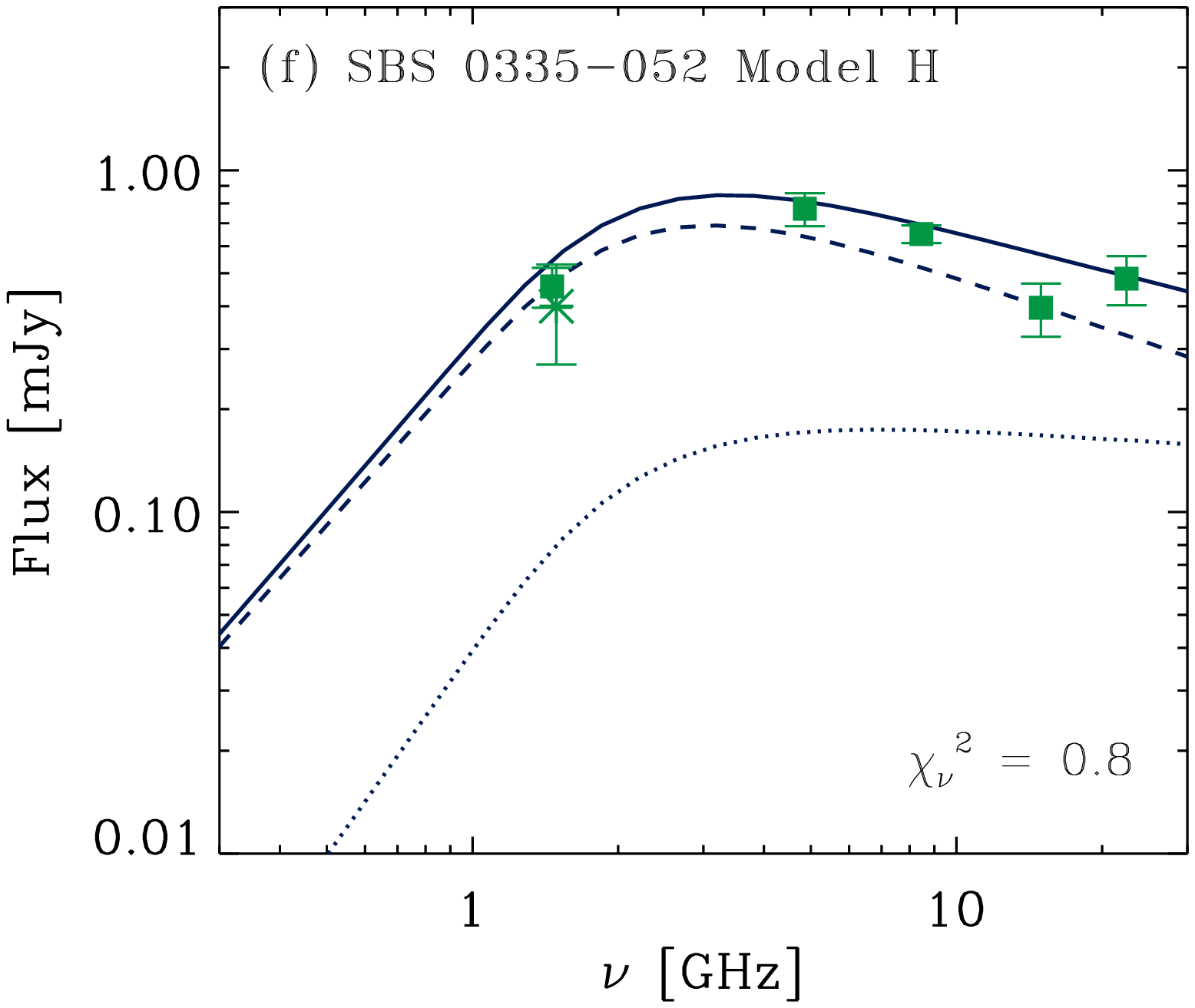}
\caption{Radio SEDs at $t=5$ Myr calculated by
(a) Model A, (b) Model B, (c) Model E, (d) Model F,
(e) Model G, and (f) Model H for the same initial
condition as Figure \ref{fig:sbs_basic}. For
Model E, we adopt the best
fit parameter of $l_{\rm nt}\tau_{\rm nt}$.
The dotted and dashed lines
represent the thermal and nonthermal components,
respectively, and the solid line shows the sum of
those two components.
The points are observational data taken from
Hunt et al.\ (\cite{hunt04a}) (squares) and
Dale et al.\ (\cite{dale01}) (cross).
The reduced $\chi^2$ value ($\chi_\nu^2$) is also shown in
each panel.
}
\label{fig:sbs}
\end{figure*}

We show the radio SEDs calculated by various models
listed in Table \ref{tab:model}, and examine whether or not the
model predictions agree with observational data. 
The radio data are shown in Figure \ref{fig:sbs}:
the squares are from H04, and the cross is from
Dale et al.\ (\cite{dale01}). The SED at $t=5$ Myr is
presented in Figure \ref{fig:sbs},
where the dotted and dashed lines represent the
nonthermal and thermal components, respectively.
The sum of those two components is shown by the
solid line in each panel. The panel (a) shows the
result of Model A, in which we adopt the
``standard'' model for the nonthermal radio emission.
In this case, the thermal emission dominates the
radio SED. The model prediction is inconsistent with
the observational SED in the following two aspects:
(i) the theoretical flux is systematically smaller
than the observed one; (ii) the flat SED, dominated
by the free-free component, is inconsistent with the
observational SED at $\nu\ga 4$ GHz,
which requires a prominent nonthermal contribution.
Indeed H04 show that the nonthermal fraction of
the radio luminosity at 5 GHz is $\sim 0.7$.
If we adopt Model B with a burst SFR instead of Model A, the
thermal component decreases, because the mean
age of the stars is older. In this case also,
however, the nonthermal radiation
of Model B falls well below the data presented in Figure
\ref{fig:sbs}.
As long as we adopt the ``standard''
nonthermal component (i.e., Models A--D), we cannot
explain the prominent nonthermal contribution.
Therefore, we exclude
Models A--D, and adopt the parameterized model for
the nonthermal component (i.e., Models E--H).

In the parameterized model of the nonthermal
component, $l_{\rm nt}\tau_{\rm nt}$ is treated
as a parameter to be determined from observations
(Eq.\ \ref{eq:nt_para}). Hence, we estimate
$l_{\rm nt}\tau_{\rm nt}$ by using the observational
data of \sbs\ taken by H04 and
Dale et al.\ (\cite{dale01}). We adopt the same
initial conditions and parameters as above, and assume
an age of $t=5$ Myr; we then 
search for the value of $l_{\rm nt}\tau_{\rm nt}$
in Models E--H which minimizes $\chi^2$
for the six data points. 
The best-fit $l_{\rm nt}\tau_{\rm nt}$ are listed in
Table \ref{tab:parameter} with $\chi^2$ and
$\chi_\nu^2$ (reduced $\chi^2$; the number of
freedom is five), and the
best-fit SEDs for Models E--H are shown in
Figures \ref{fig:sbs}c--f, respectively.

  \begin{table}
     \caption[]{Best Fit Parameters for the Nonthermal
                Component (\sbs)}
        \label{tab:parameter}
\begin{tabular}{@{}cccc@{}}\hline
Model & $l_{\rm nt}\tau_{\rm nt}(5~{\rm GHz})$ & $\chi^2$
& $\chi_\nu^2$ \\
& (W Hz$^{-1}$ yr) & & \\
\hline
E & $7.6\times 10^{22}$ & 6.8 & 1.4 \\
F & $22\times 10^{22}$ & 34 & 6.8 \\
G & $5.8\times 10^{22}$ & 11 & 2.2 \\
H & $19\times 10^{22}$ & 4.0 & 0.8 \\
\hline
\end{tabular}
  \end{table}

The burst models require higher $l_{\rm nt}\tau_{\rm nt}$.
In the burst models, all the stars form at
the beginning, so that the mean stellar
age is older. Thus, the thermal emission of
the burst models is fainter than that of the
continuous star formation models.
This is why we require a higher
nonthermal time-luminosity integral of a SN II in the burst
SFR models (Models F and H) than in the continuous SFR models
(E and G) to explain the observed radio emission.
With the nonthermal component comparable to or larger than
the thermal component, the points at 4.86, 8.46, and
22.5 GHz are well reproduced. The spectral slope of
the nonthermal component is also consistent with
the observations. Thus nonthermal radio
emission comparable to or stronger than the
thermal radio emission is required
even in this young ($<5$ Myr) galaxy.
In general, the fits are satisfactory for
Models E--H except for the point at
14.9 GHz.
The fractions of nonthermal
emission are 0.44, 0.79, 0.37, and 0.79 at
5 GHz in Models E, F, G, and H, respectively.
The fractions in Models F and H are more consistent with 
the value of 0.7 at 5\,GHz inferred from the spectral 
decomposition by H04.

The nonthermal feature of the radio spectrum
is inconsistent with an age of $t<3.5$ Myr, because of the
time required for the onset of SNe. On the other hand,
as long as we adopt an age between 4 Myr and 7 Myr,
the difference in the derived
$l_{\rm nt}\tau_{\rm nt}$ is within a factor of $\sim 2$
(i.e., within the uncertainty among Models E--H.)
Thus, we conclude that $l_{\rm nt}\tau_{\rm nt}$ derived
here is robust against the age uncertainty.

The nonthermal radiative energy emitted at $\nu =5$\,GHz
over the entire lifetime of a SNR is predicted to be
5.8--$22\times 10^{22}$ W Hz$^{-1}$ yr
(Table \ref{tab:parameter}). This is more than
an order of magnitude larger than the value used in the
``standard'' model, in which the nonthermal
radio energy emitted by a SNR is estimated based on
the observed $\Sigma -D$ (surface luminosity vs.\
radius) relation and the theoretical adiabatic lifetime.
That assumption leads
to the total radio energy per unit frequency
as (assuming $\alpha =-0.5$)
\begin{eqnarray}
E_{\rm nt}^{\rm s}(\nu ) & = & 1.1\times 10^{29}
E_{\rm 51}^{-1/17}\left(
\frac{n_{\rm H}}{1~{\rm cm}^{-3}}\right)^{-2/17}
\nonumber \\
& \times & \left(\frac{\nu}{5~{\rm GHz}}
\right)^{-0.5}
~{\rm J~{\rm Hz}^{-1}}\, ,\label{eq:Es}
\end{eqnarray}
while our new result suggests it to be
\begin{eqnarray}
E_{\rm nt}^{\rm p}(\nu ) & = &
(1.8\mbox{--}7.1)\times 10^{30}\left(
\frac{\nu}{5~{\rm GHz}}\right)^{-0.5}
~{\rm J~{\rm Hz}^{-1}}\, .
\end{eqnarray}
Integrating those quantities from 0.1 to 100 GHz,
we obtain the total radio energy per SNR as
\begin{eqnarray}
{\cal E}^{\rm s}(0.1\mbox{--}100~{\rm GHz}) & = &
4.7\times 10^{39}E_{\rm 51}^{-1/17} \nonumber \\
& & \times \left(
\frac{n_{\rm H}}{1~{\rm cm}^{-3}}\right)^{-2/17}
~{\rm J}\,\label{eq:calEs}
\end{eqnarray}
for the ``standard'' case, and
\begin{eqnarray}
{\cal E}^{\rm p}(0.1-100~{\rm GHz}) & = &
(8\mbox{--}31)\times 10^{40}~{\rm J}\,
\end{eqnarray}
for the ``parameterized'' case.

\afterpage{\clearpage}

\subsection{\izw}\label{subsec:izw}

We now apply our model to another ``template'' of nearby
young BCDs, \izw, whose nebular oxygen abundance is low
(12$+$log(O/H)=7.2, Skillman \& Kennicutt \cite{skillman93}). 
The star-forming region of \izw\ is more diffuse
than that of \sbs\ (e.g.,
Hunt et al.\ \cite{hunt-cozumel}).
Thus, the different gas density of \izw\ from
\sbs\ may provide us with independent
information on the density dependence of radio
emission.

For the initial density and the radius, we assume
$n_{\rm H0}=100~{\rm cm}^{-3}$ and $r_{\rm SF}=100~{\rm pc}$,
consistently with optical spectra and images 
(Skillman \& Kennicutt \cite{skillman93}; Hunt et al.\ \cite{hunt-cozumel};
Hirashita \& Hunt \cite{hirashita04}).
The resulting gas mass is
$M_{\rm gas}=1.4\times 10^7~M_\odot$ with
$\tff =4.4\times 10^6$ yr; the gas mass is comparable to the
observationally determined values by
van Zee et al.\ (\cite{vanzee98}) ($2.6\times 10^7~M_\odot$
for their H\,{\sc i}-A component)
and Lequeux \& Viallefond (\cite{lequeux80})
($2.1\times 10^7~M_\odot$ for their component 1).

We adopt an age of 10--15\,Myr for the dominant burst in \izw.
This is consistent with the results of our model of infrared
dust emission
(Hirashita \& Hunt \cite{hirashita04}), which gives an
age of 10--15 Myr and consistent with the observed metallicity and 
dust abundance.
Takeuchi et al.\ (\cite{takeuchi03}) show that
the dust mass derived by Cannon et al.\ (\cite{cannon02})
is consistent with an age of 10--30 Myr, which is also
consistent with the observational upper limits of
infrared flux. Recchi et al.\ (\cite{recchi04})
(and references therein)
argue that stars older than 0.5--1 Gyr,
if any, do not produce a significant
contribution to the metal budget of
\izw, but intermediate
age stars with an age of a few hundred Myr
may be required from studies of the chemical
abundances. An analysis of the color-magnitude
diagram by Aloisi et al.\ (\cite{aloisi99})
also suggests the existence of intermediate-age
populations. Hunt et al.\ (\cite{hti03})
derive an age of $<500$ Myr for the
oldest stellar population (but see
\"{O}stlin \cite{ostlin00}) and $\la 15$ Myr
for the youngest one in the main body.

\begin{figure*}
\includegraphics[width=8cm]{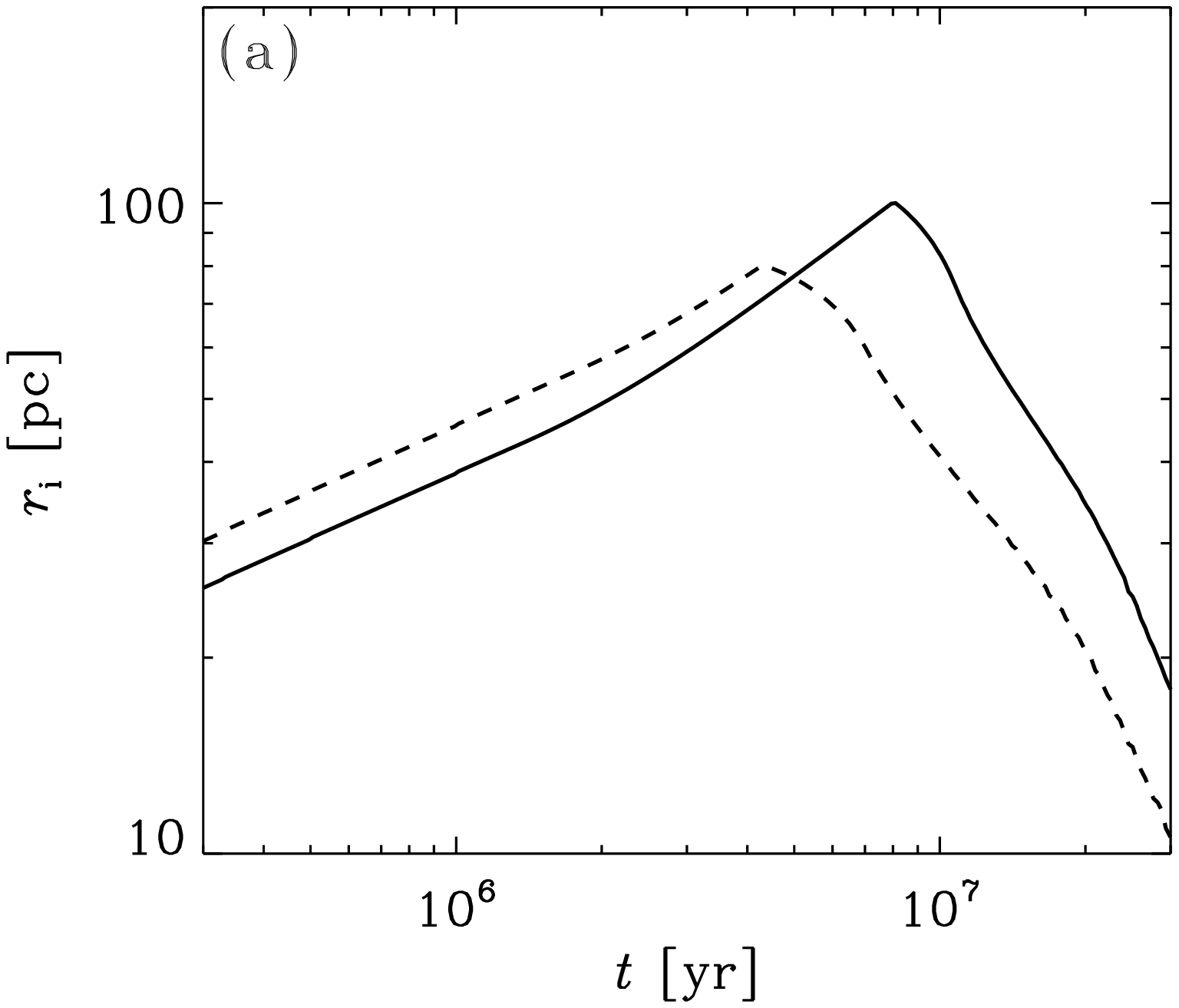}
\includegraphics[width=8cm]{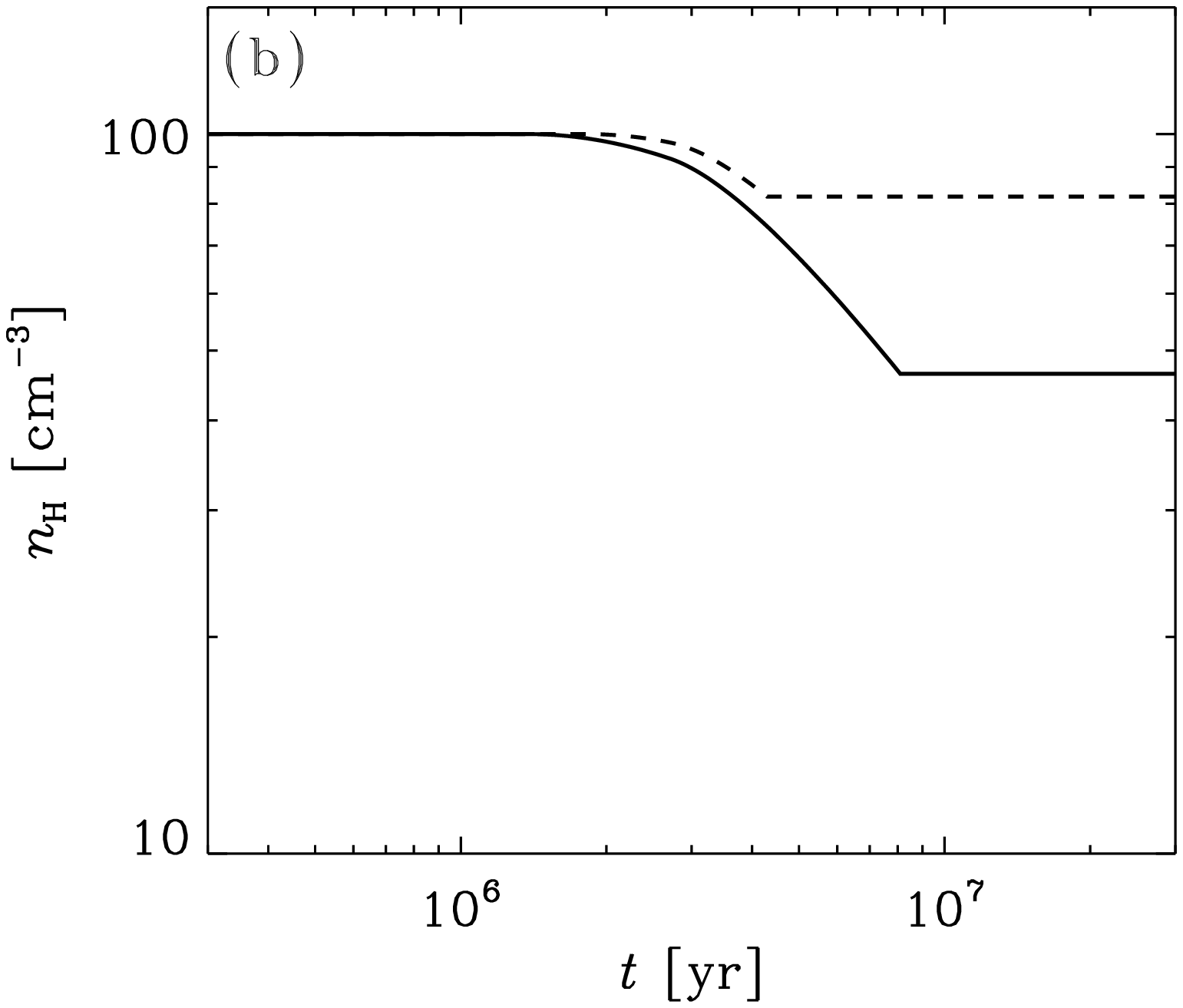}\\
\includegraphics[width=8cm]{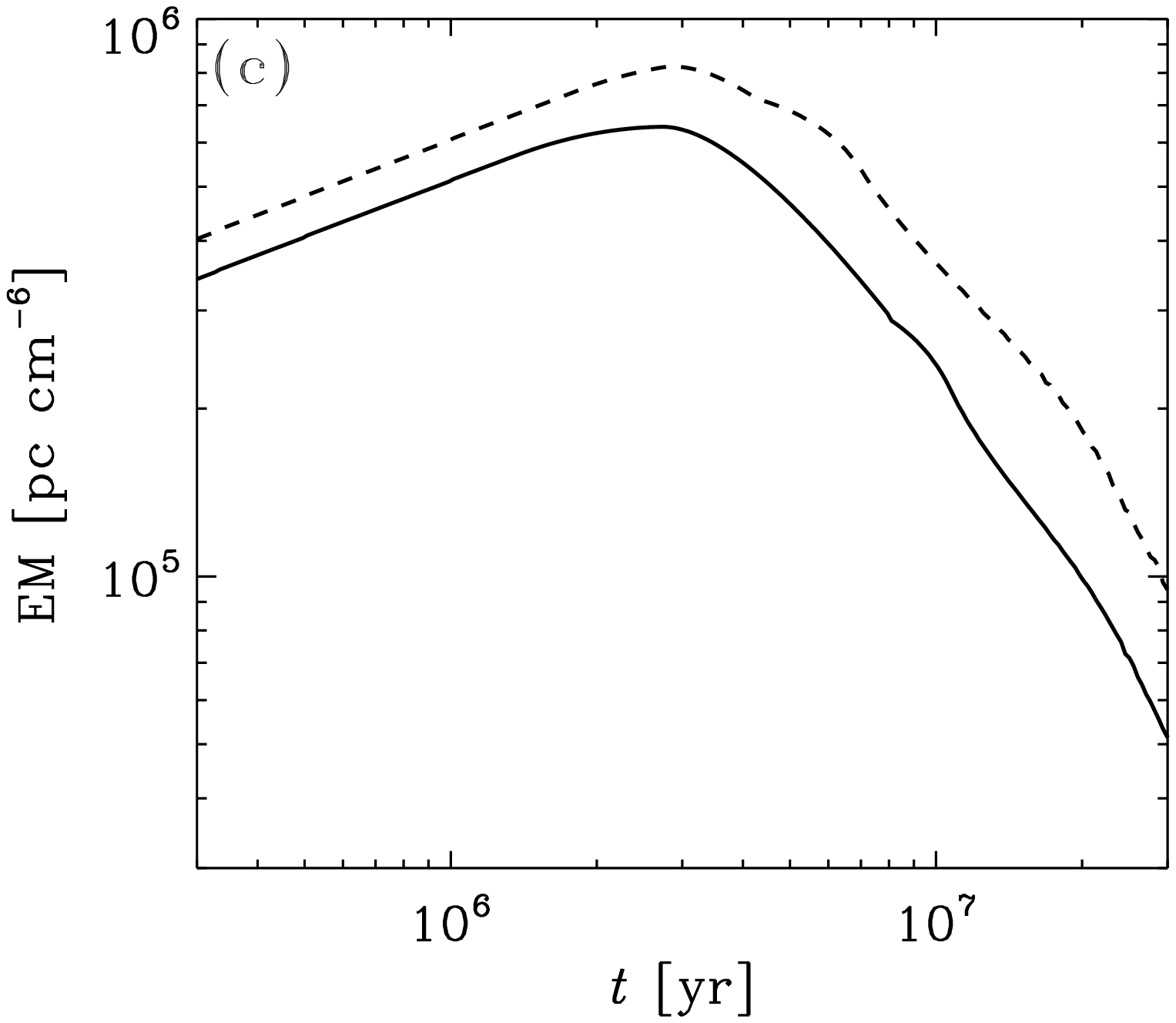}
\includegraphics[width=8cm]{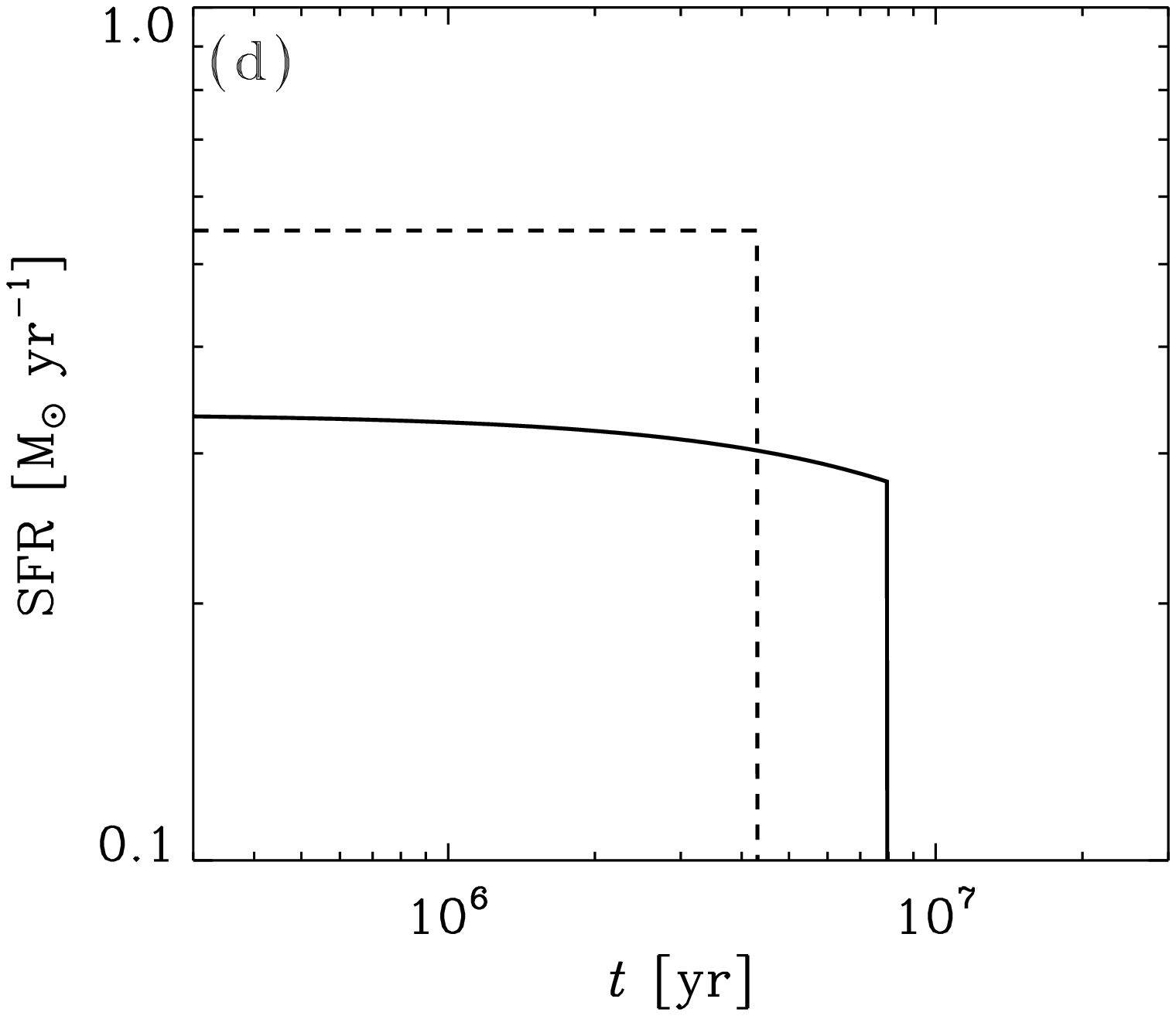}
\caption{Same as Fig.\ \ref{fig:sbs_basic}
but for the \izw\ ``passive'' model.}
\label{fig:izw_basic}
\end{figure*}

First, we evaluate a continuous SFR.
The time evolution of $r_{\rm i}$, $n_{\rm H}$,
EM, and SFR is shown in
Figures \ref{fig:izw_basic}a--d. 
In the first 1.4 Myr,
the expansion speed of the ionizing front is faster
than the sound speed because of small gas density.
Thus, the dynamical response of the system begins
later compared to \sbs. Moreover, because of the
low density, 
the emission measure remains smaller than in \sbs.
The star formation
stops at $t=7.9$ Myr because the entire star-forming
region is ionized. The total stellar mass formed is
$2.4\times 10^6~M_\odot$.

In the burst SFR, the SFR is constant up to
$t_{\rm dyn}=4.4$ Myr, a little shorter than the
above duration (7.9 Myr). Thus, the expansion of
the ionized region stops earlier and $n_{\rm H}$
remains higher. However, the decline of $r_{\rm i}$
occurs earlier because of the shorter duration,
and the resulting EM is roughly similar to that
for the continuous SFR.
As a result, there is little difference in
the evolution of emission measure between the
continuous SFR and the burst SFR in a low-density regime.

We also calculate radio SEDs with various models listed
in Table \ref{tab:model}. As mentioned above, we adopt an
age of 10--20 Myr.
Because the (continuous mode of) star formation stops at
$t=7.9$ Myr, the radio emission
decreases between $t=10$ and 20 Myr. At 10 Myr, the
thermal component is too high to be consistent with the 
observations reported by
Hunt et al.\ (\cite{hunt05}).
The age should be older than 12 Myr to be consistent
with the data. At 15 Myr, the thermal
component is well below the observations.
Thus, we examine the two ages,
$t=12$ Myr and 15 Myr, as high and low limiting luminosities for
the thermal component.
The emission measures at $t=12$ Myr and 15 Myr are
$2.4\times 10^5$ pc cm$^{-6}$ and
$1.4\times 10^5$ pc cm$^{-6}$, respectively.
Those are larger than the EM estimated by
Hunt et al.\ (\cite{hunt05})
(${\rm EM}\sim 10^4$ pc cm$^{-6}$), but smaller
than their extreme case
(${\rm EM}\sim 10^7$ pc cm$^{-6}$).
If we adopt an age of $\ga 20$ Myr, the emission
measure becomes too small to be consistent with
observations.

\begin{figure*}
\includegraphics[width=8cm]{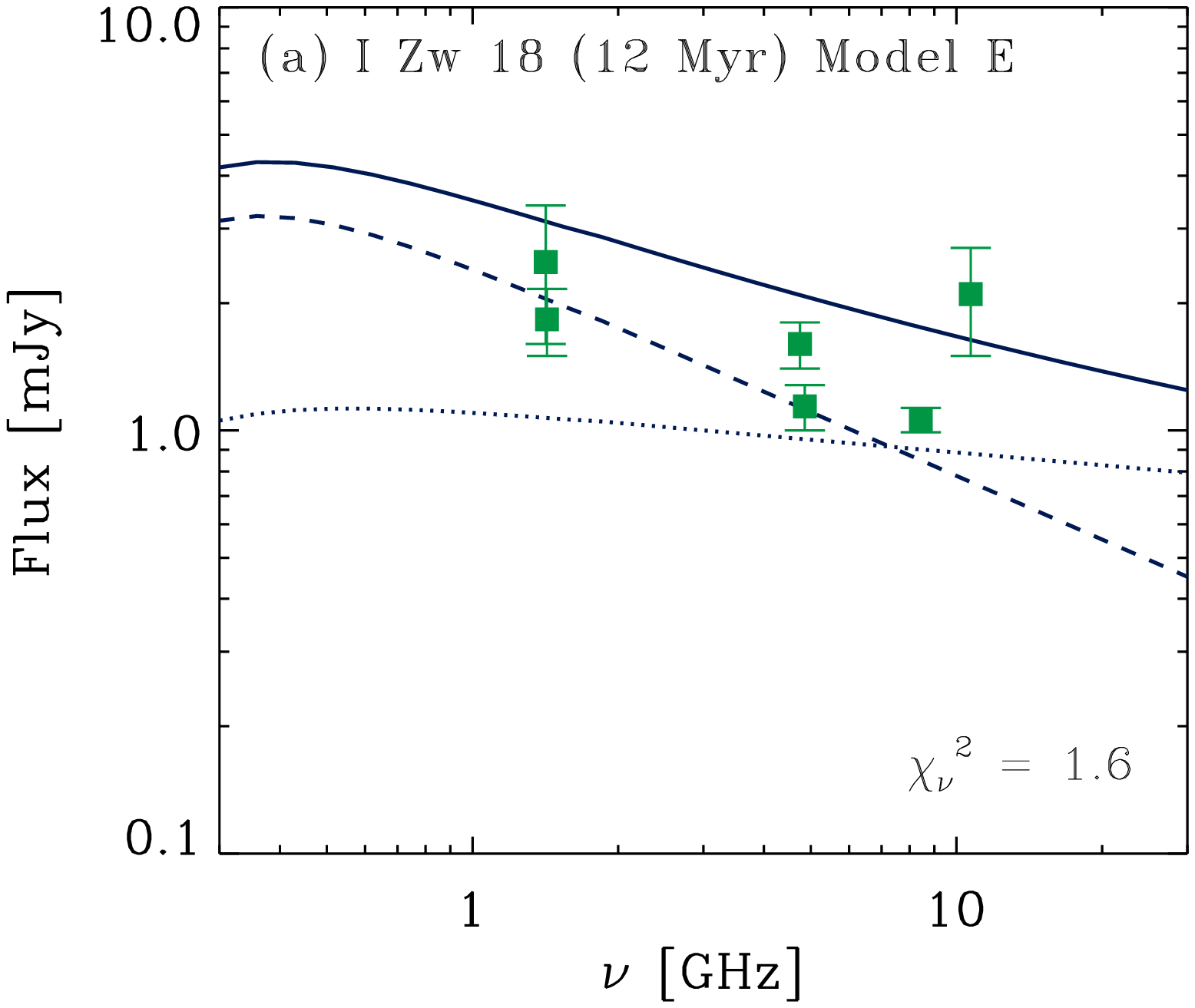}
\includegraphics[width=8cm]{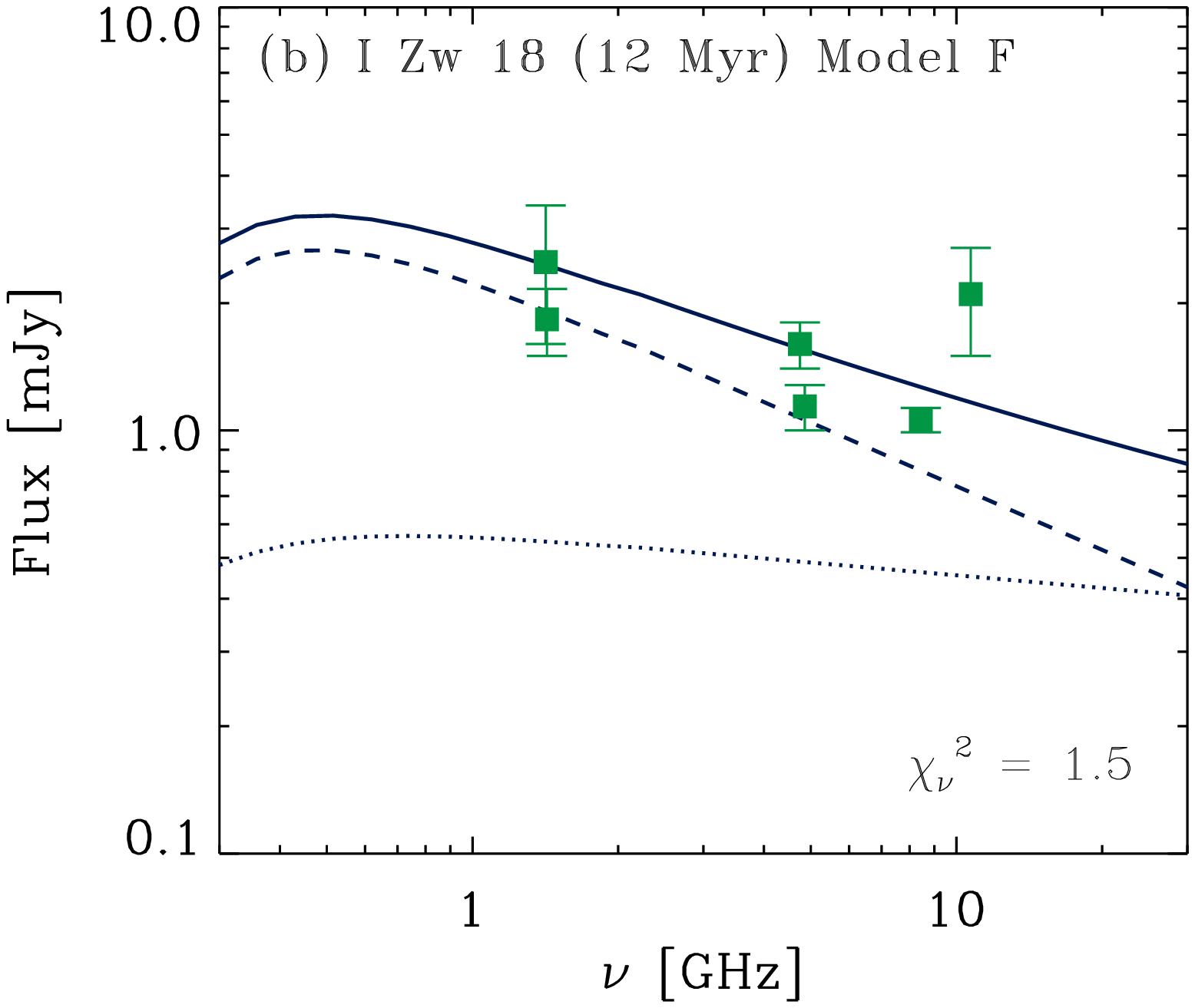}\\
\includegraphics[width=8cm]{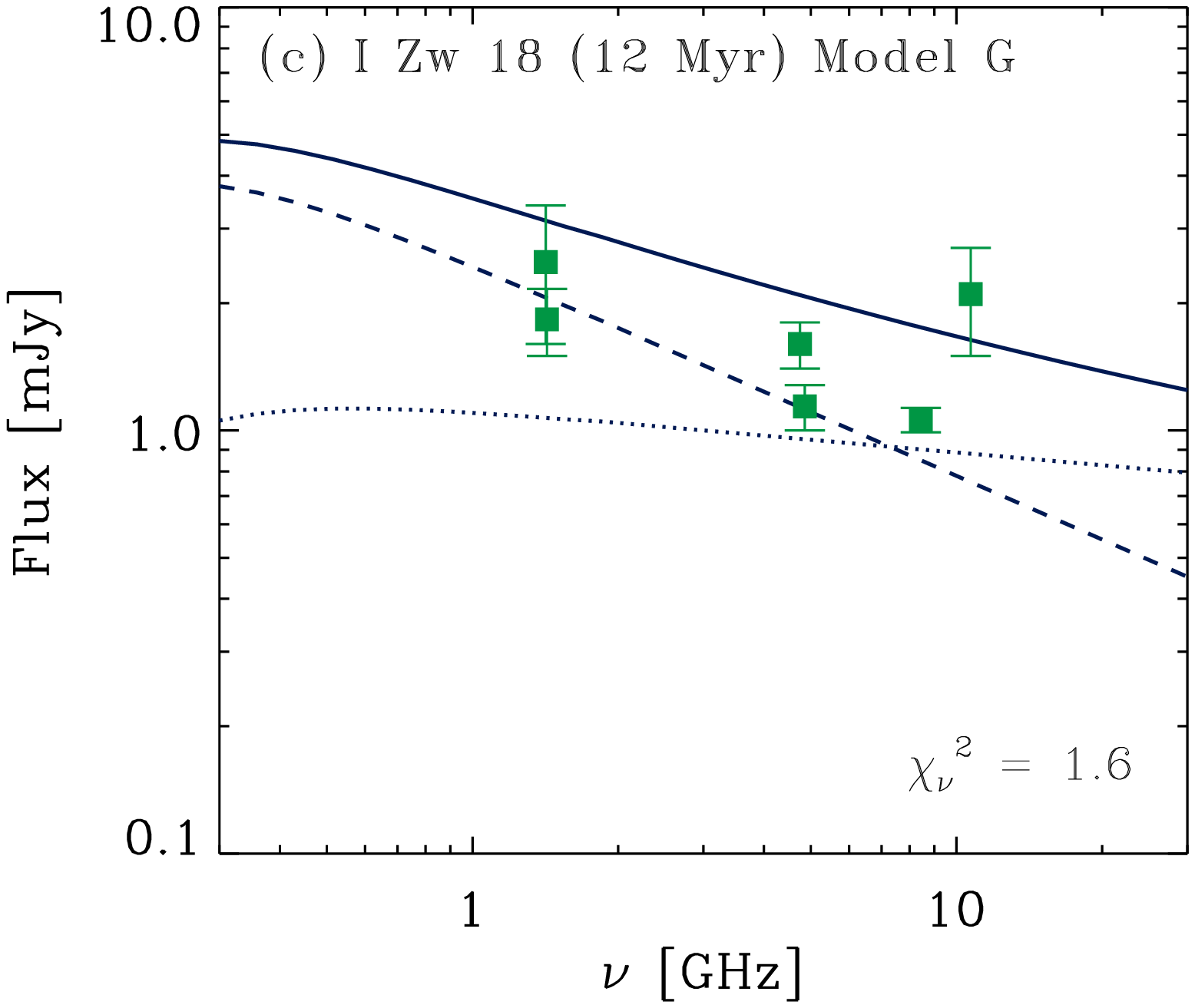}
\includegraphics[width=8cm]{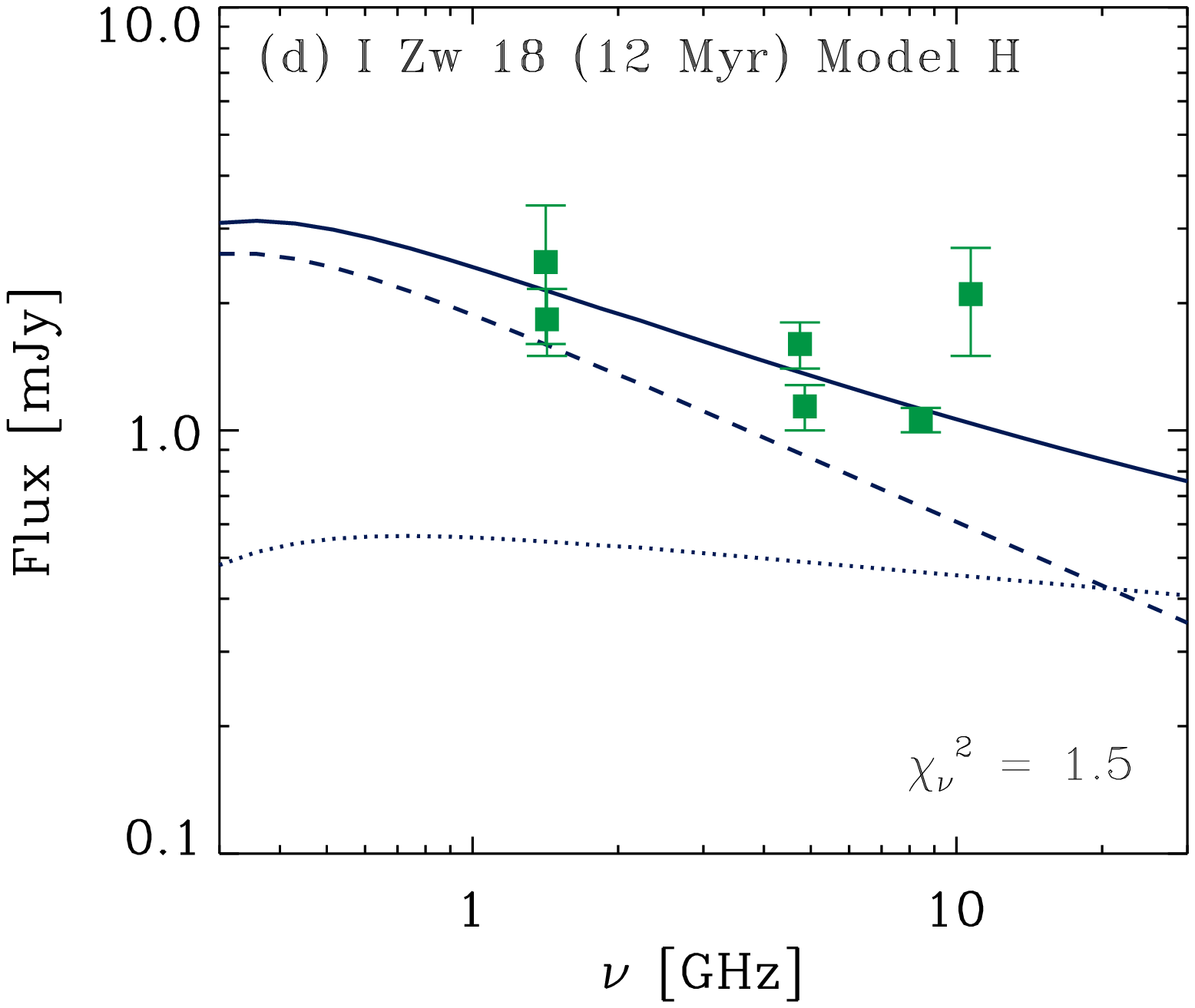}
\caption{Radio SED at $t=12$ Myr calculated for
\izw\ by using Models E, F, G, and H
(panels {\bf a)}, {\bf b)}, {\bf c)}, and
{\bf d)}, respectively).
The dotted and dashed lines
represent the thermal and nonthermal components,
respectively, and the solid line shows the sum of
those two components.
The squares are observational data taken from
Hunt et al.\ (\cite{hunt05}) and
references therein. The reduced $\chi^2$ value
($\chi_\nu^2$) is also shown in
each panel.
}
\label{fig:izw}
\end{figure*}

\begin{figure*}
\includegraphics[width=8cm]{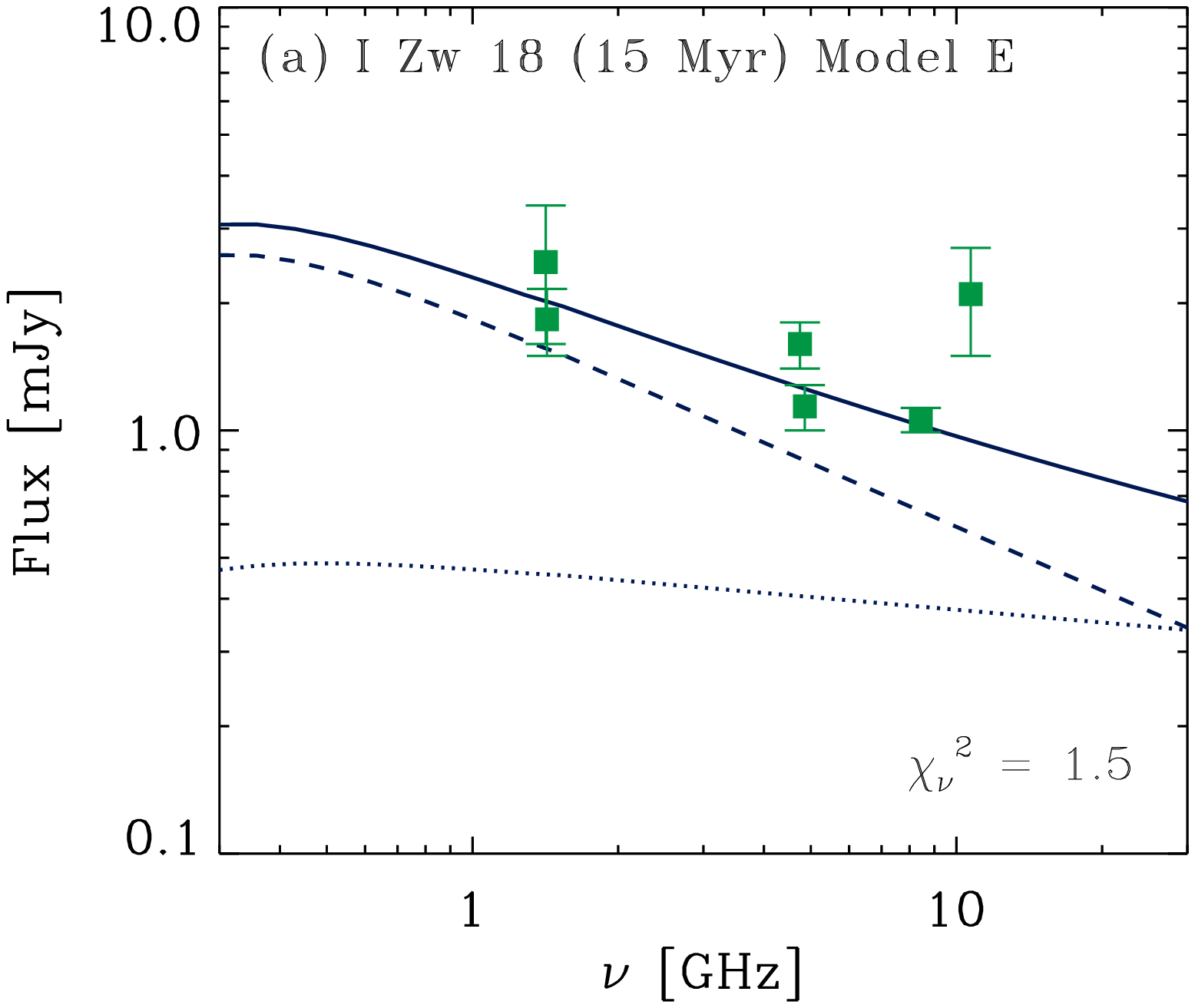}
\includegraphics[width=8cm]{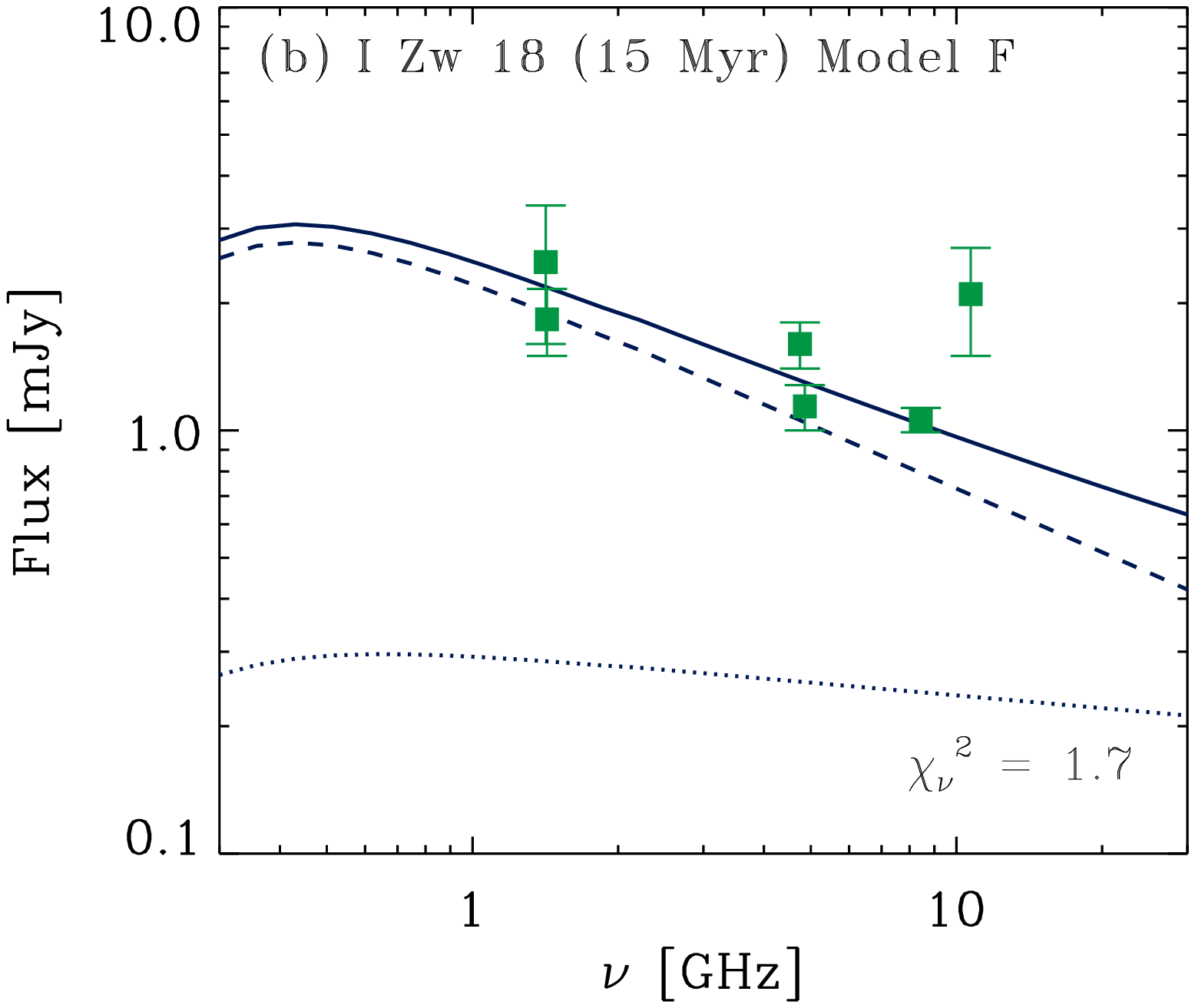}\\
\includegraphics[width=8cm]{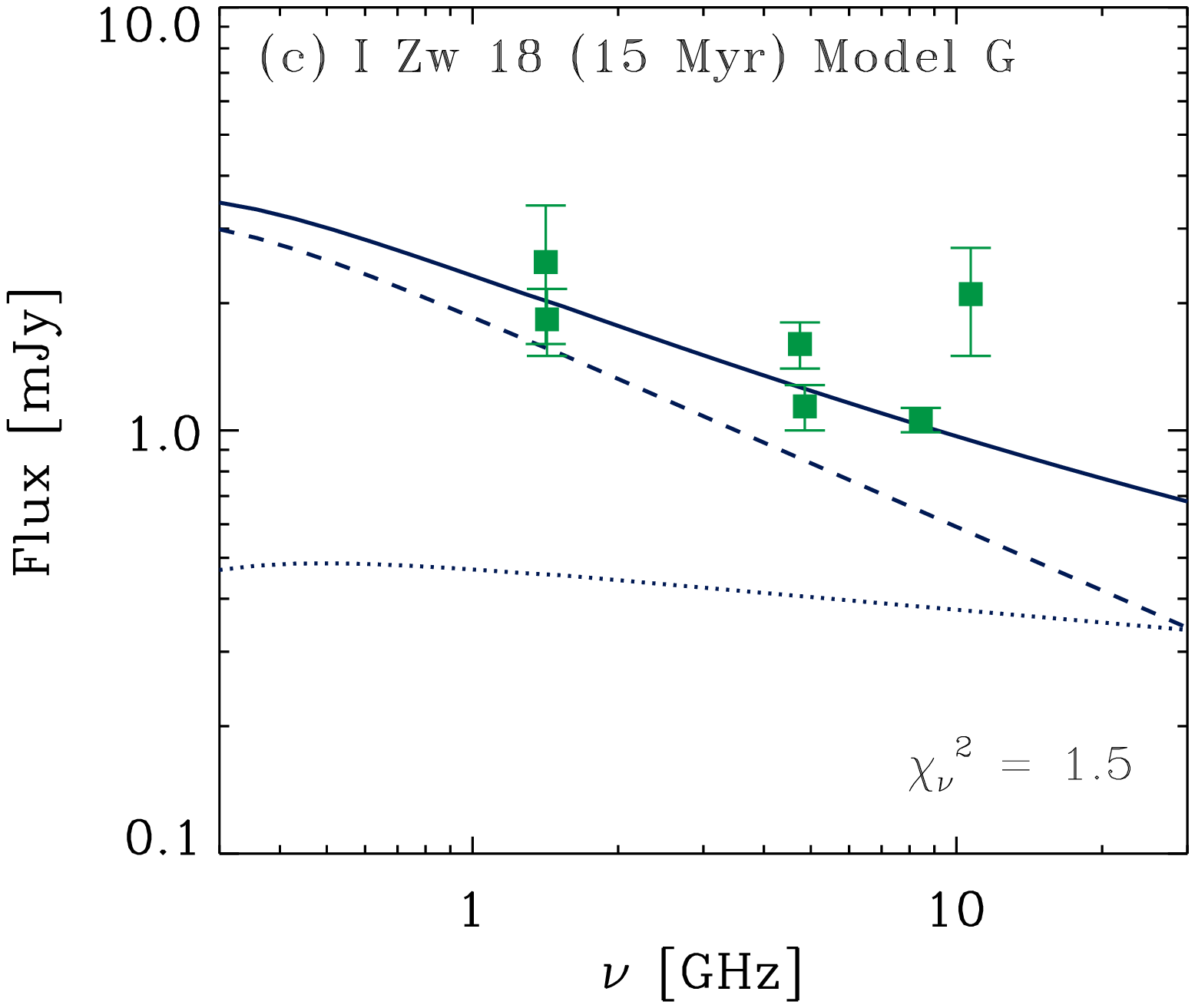}
\includegraphics[width=8cm]{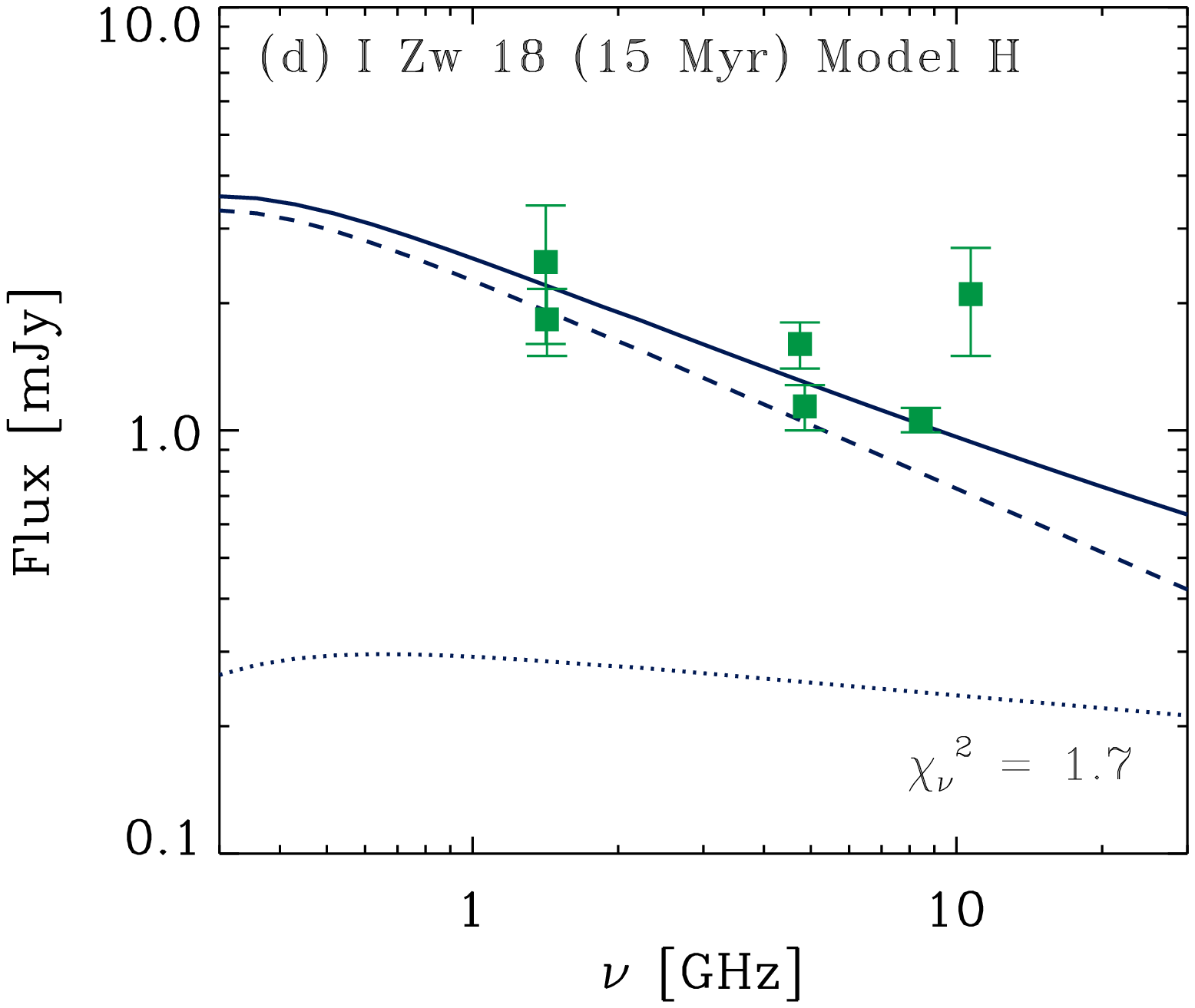}
\caption{Same as Fig.\ \ref{fig:izw} but for
$t=15$ Myr.
}
\label{fig:izw15}
\end{figure*}

Now we examine the radio SED at $t=12$ Myr, but
consider only the p-models for the nonthermal
component, since the s-models are rejected after the
investigation of \sbs. In fact, the s-models are
also unacceptable for \izw.
As in Sect.\ \ref{subsec:sbs}, 
the best-fit nonthermal time-integrated luminosity (fluence)
per SN II at 5 GHz ($l_{\rm nt}\tau_{\rm nt}$) is
sought by minimizing $\chi^2$ for the six data
points adopted from Hunt et al.\ (\cite{hunt05}) and
references therein. The data are shown in
Figs.\ \ref{fig:izw} and \ref{fig:izw15}:
the lower three data points are from
the VLA (Hunt et al.\ \cite{hunt05}), while the upper
three are from single-dish radio telescopes (Klein et al.\ \cite{klein91}).
Although the discrepancy between those two data sets may
arise from the sensitivity for the diffuse component,
the discrepancy is only $\sim 40$\% at $\nu\sim 5$ GHz.
Thus, the following conclusions are not changed by the
uncertainty in the diffuse component.

\begin{figure*}
\centering
\includegraphics[width=8cm]{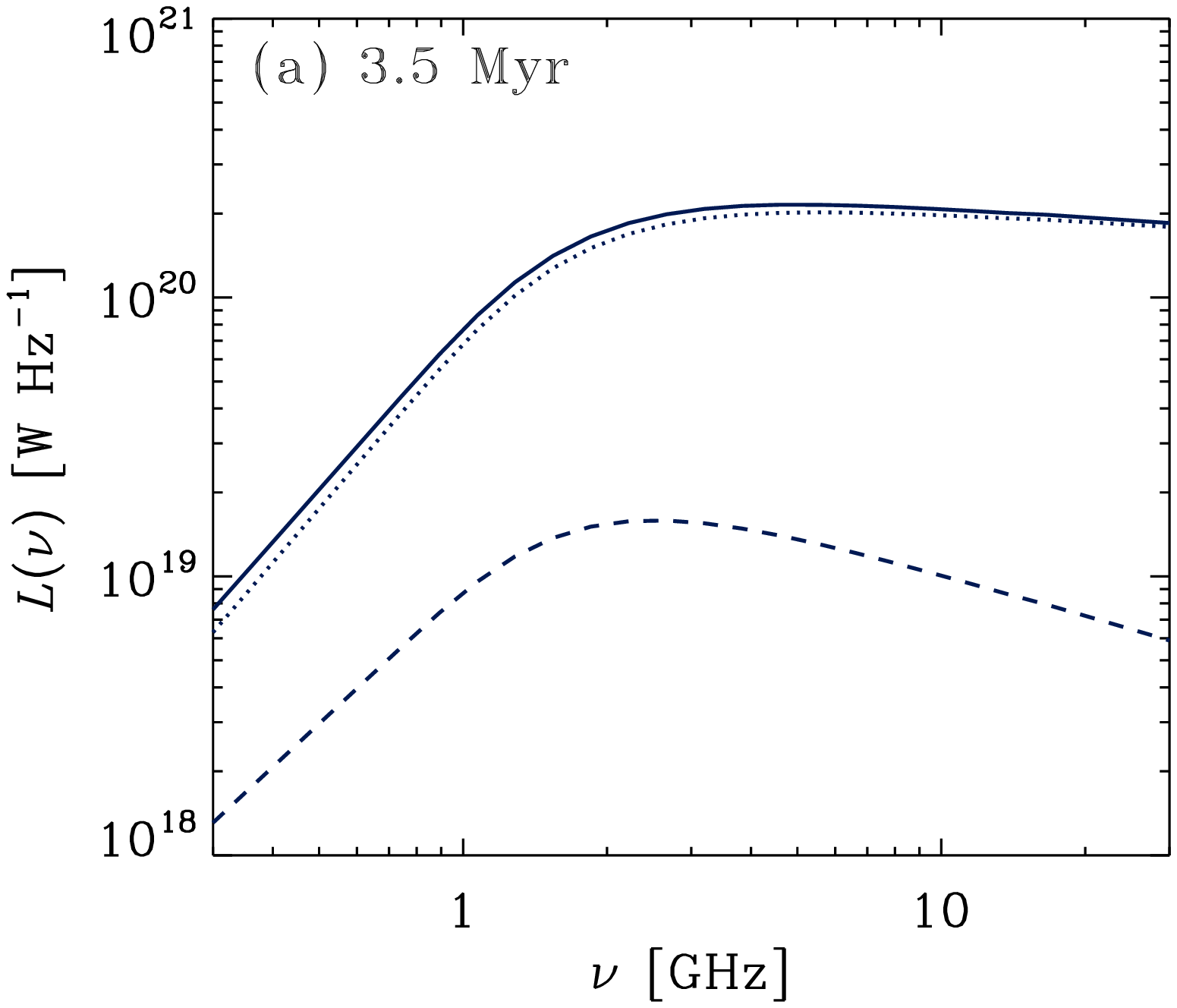}
\includegraphics[width=8cm]{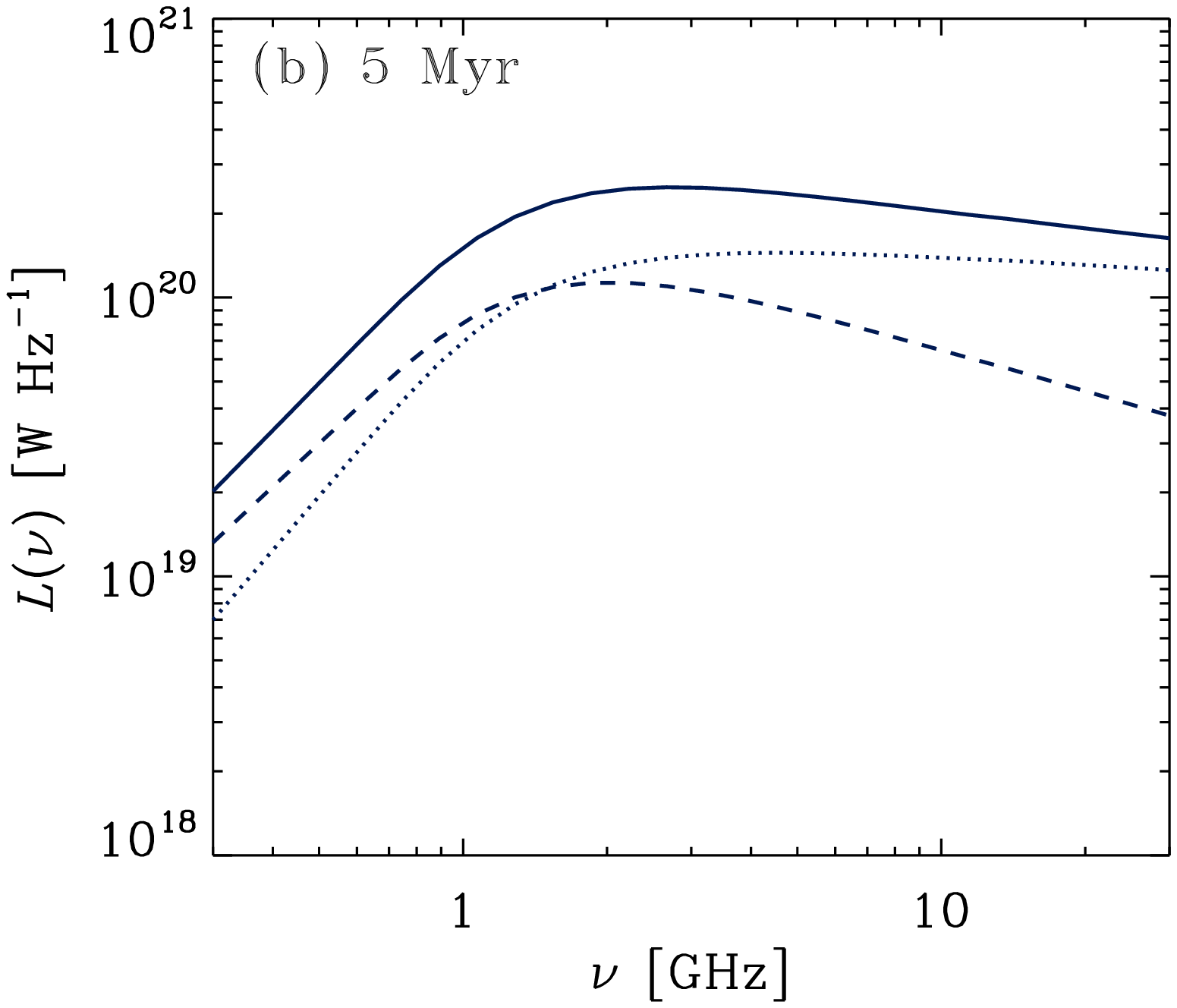}\\
\includegraphics[width=8cm]{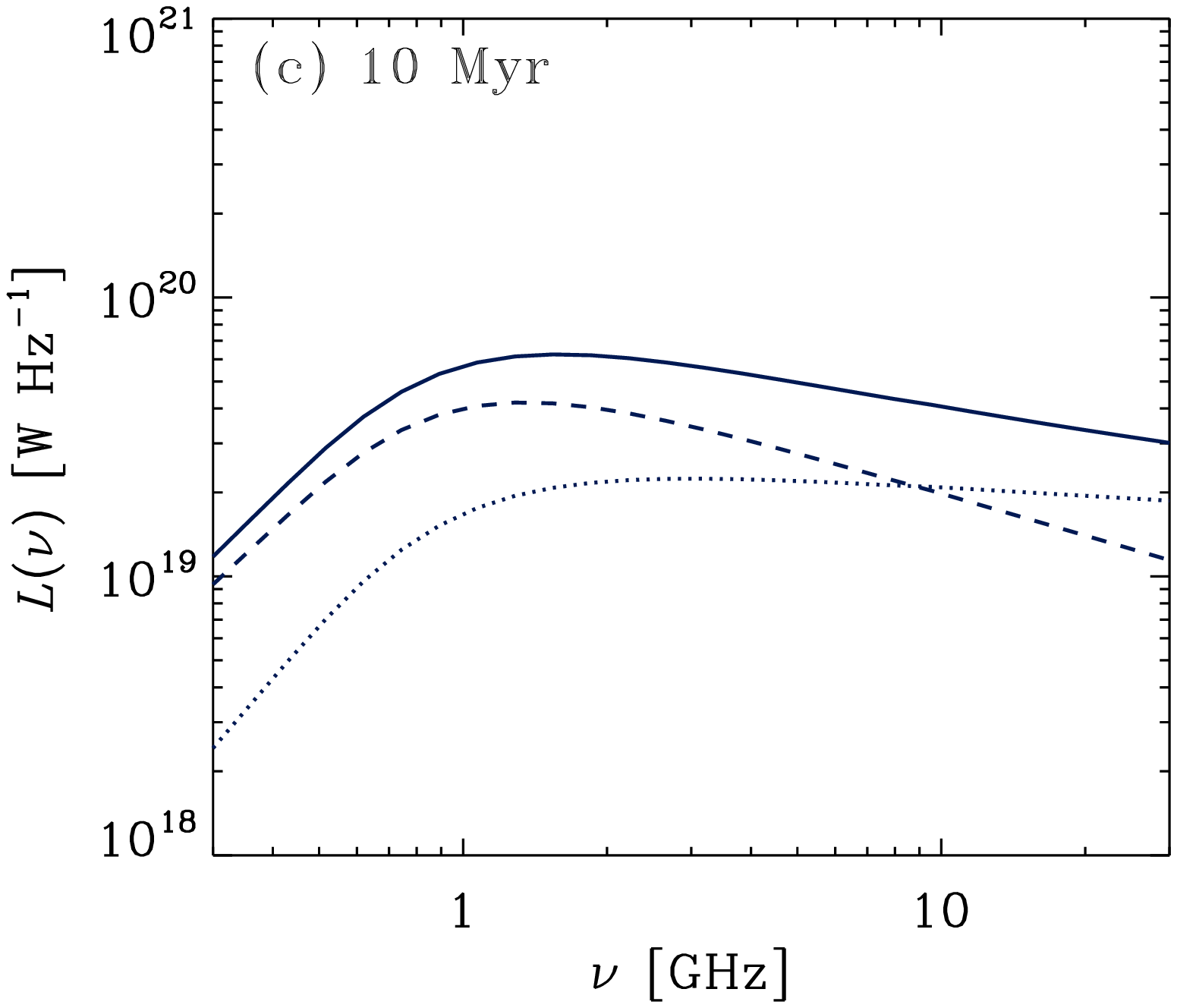}
\includegraphics[width=8cm]{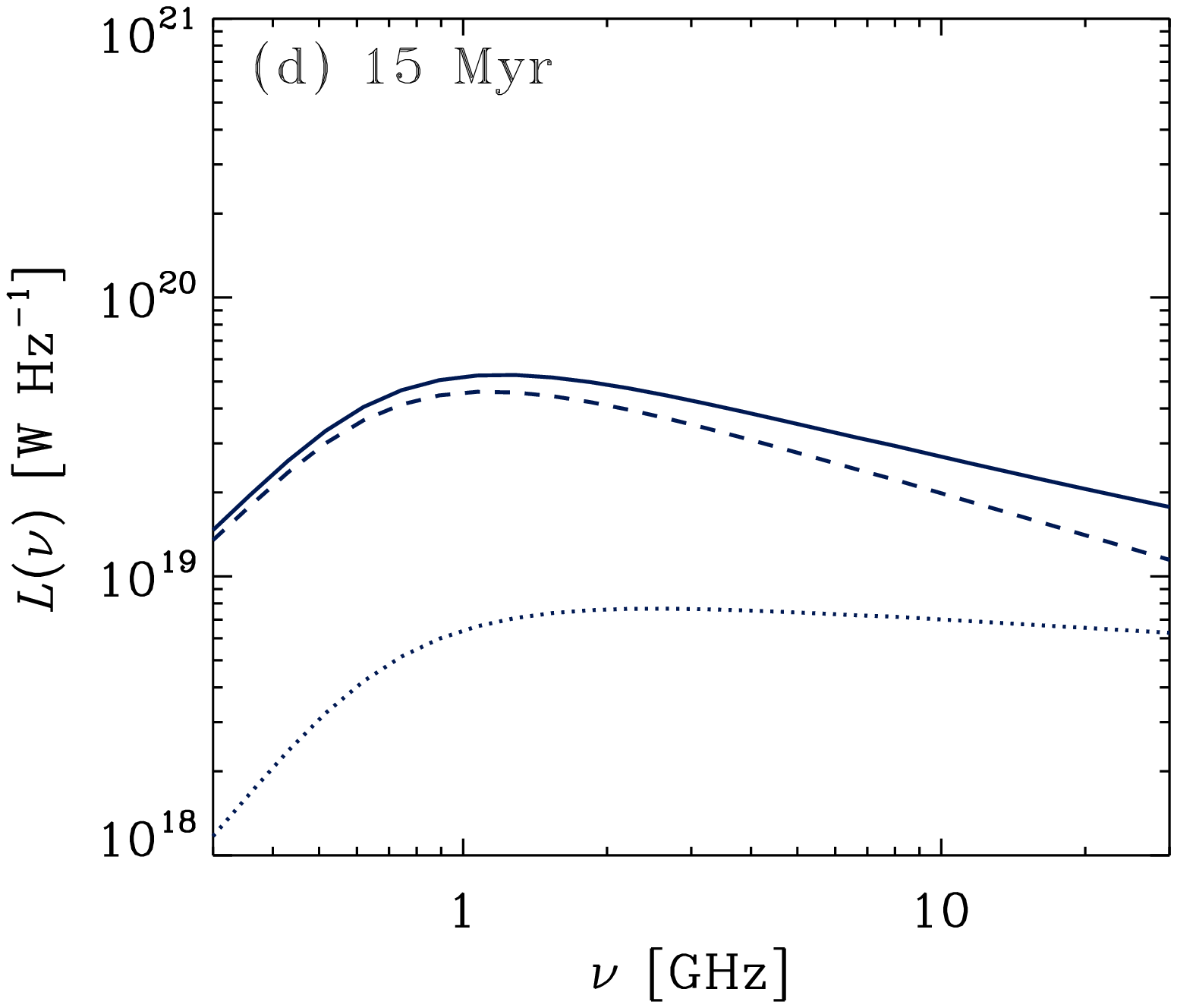}
\caption{Time evolution of an ``active'' radio SED.
The same parameters as used to model \sbs\
are adopted as a representative case for the
``active'' mode. Panels {\bf a)}, {\bf b)},
{\bf c)}, and {\bf d)} represent the SEDs at
3.5, 5, 10, and 15, respectively.
The dotted and dashed lines
represent the thermal and nonthermal components,
respectively, and the solid line shows the sum of
those two components.}
\label{fig:ev_active}
\end{figure*}

We include all the six data points to obtain the best-fit values
for $l_{\rm nt}\tau_{\rm nt}$, which are shown in
Table \ref{tab:param_izw12} for $t=12$ Myr, and
in Table \ref{tab:param_izw15} for $t=15$ Myr. Applying
the values reported in those tables, we examine the
radio SEDs. The results are shown in
Figs.\ \ref{fig:izw} and \ref{fig:izw15} for
$t=12$ and 15 Myr, respectively. Those figures confirm
that the nonthermal luminosity derived from the
parameters in Tables \ref{tab:param_izw12} and
\ref{tab:param_izw15} is consistent with the
data points. Moreover, the spectral index
at $\nu\la 5$ GHz is reproduced by introducing a
significant contribution from the nonthermal
component.

  \begin{table}
     \caption[]{Best Fit Parameters for the Nonthermal
                Component (\izw\ at $t=12$ Myr)}
        \label{tab:param_izw12}
\begin{tabular}{@{}cccc@{}}\hline
Model & $l_{\rm nt}\tau_{\rm nt}(5~{\rm GHz})$ & $\chi^2$
& $\chi_\nu^2$ \\
& (W Hz$^{-1}$ yr) & & \\
\hline
E & $1.1\times 10^{22}$ & 8.2 & 1.6 \\
F & $2.5\times 10^{22}$ & 7.5 & 1.5 \\
G & $1.1\times 10^{22}$ & 8.2 & 1.6 \\
H & $2.5\times 10^{22}$ & 7.6 & 1.5 \\
\hline
\end{tabular}
  \end{table}

\begin{table}
     \caption[]{Best Fit Parameters for the Nonthermal
                Component (\izw\ at $t=15$ Myr)}
        \label{tab:param_izw15}
\begin{tabular}{@{}cccc@{}}\hline
Model & $l_{\rm nt}\tau_{\rm nt}(5~{\rm GHz})$ & $\chi^2$
& $\chi_\nu^2$ \\
& (W Hz$^{-1}$ yr) & & \\
\hline
E & $2.4\times 10^{22}$ & 7.7 & 1.5 \\
F & $3.0\times 10^{22}$ & 8.5 & 1.7 \\
G & $2.4\times 10^{22}$ & 7.7 & 1.5 \\
H & $3.0\times 10^{22}$ & 8.6 & 1.7 \\
\hline
\end{tabular}
\end{table}

The spectral decomposition by Hunt et al.\ (\cite{hunt05})
gives a nonthermal fraction of 0.59 at 5\,GHz, corresponding
to a thermal flux of 0.54 mJy.
This is approximately realized for Models F and H
shown in Panels {\bf b)} and
{\bf d)} of Fig.\ \ref{fig:izw}, and for Models E
and G shown in Panels
{\bf a)} and {\bf c)} of Fig.\ \ref{fig:izw15}.
However, other solutions with a thermal component
of 0.3--1 mJy are also permitted.
The younger age of 12\,Myr gives
0.20 $M_\odot~{\rm yr}^{-1}$ for the averaged SFR,
while 15\,Myr gives 0.16 $M_\odot~{\rm yr}^{-1}$.
The massive SFR is estimated as
SFR$_{>5~M_\odot}\simeq 0.036$ and 0.029 $M_\odot$ yr$^{-1}$,
respectively. Those values are 1.5--2 times larger
than the estimate of Hunt et al.\ (\cite{hunt05}).
If we assume a smaller SFR according to Hunt et al.\
(\cite{hunt05}), $l_{\rm nt}\tau_{\rm nt}$ required to
fit the observational SED becomes 1.5--2 times smaller.

Cannon et al.\ (\cite{cannon05}) also provide
fluxes of \izw\ at $\nu =1.4$ GHz and $\nu =8.4$ GHz; their
fluxes are lower than the values given by Hunt
et al.\ (\cite{hunt05}), but it is possible that
the discrepancy comes from the sensitivity to the
diffuse component. However, they also derive a spectral
index that indicates a prominent nonthermal contribution.
Consequently, our conclusion that a strong contribution from
the nonthermal component is required should be robust 
(and $l_{\rm nt}\tau_{\rm nt}$ derived by us should be relatively secure).

We have derived values of $l_{\rm nt}\tau_{\rm nt}$ for
\izw\ between
$1.1\times 10^{22}$ W Hz$^{-1}$ yr and
$2.5\times 10^{22}$ W Hz$^{-1}$ yr at 5 GHz.
Since the radio spectrum is dominated by the
nonthermal component already at 15 Myr, 
those derived values should be robust for older ages,
because the thermal component steeply declines thereafter.
On the other hand, if we adopt an age smaller
than 10 Myr, the fraction of the thermal component
is too large to be consistent with observations.
Thus, even though there is uncertainty in the age of
\izw, we can safely conclude that
$l_{\rm nt}\tau_{\rm nt}$ of
\izw\ is two to five times smaller than that of \sbs.
The typical densities of ionized
regions are $\sim 10^3$ cm$^{-3}$ and
$\sim 10^2$ cm$^{-3}$ in \sbs\ and \izw,
respectively. 
We propose in Sect.\ \ref{sec:mag} a scaling relation
$l_{\rm nt}\tau_{\rm nt}\propto n^{0.35\mbox{--}0.73}$.
Since $l_{\rm nt}\tau_{\rm nt}=5.8\mbox{--}22\times
10^{22}$ W Hz$^{-1}$ yr has been obtained for \sbs,
the above density scaling relation predicts that
$l_{\rm nt}\tau_{\rm nt}\sim 1.1-9.9\times 10^{22}$
W Hz$^{-1}$ yr. The above values derived for \izw\
are in this range.

For \izw, we obtain the following value for the
nonthermal radiative energy over the entire
lifetime of a SNR:
\begin{eqnarray}
E_{\rm nt}^{\rm s}(\nu )=3.4\mbox{--}9.4\times 10^{29}
\left(
\frac{\nu}{5~{\rm GHz}}\right)~{\rm J~Hz}^{-1}\, .
\end{eqnarray}
As for \sbs,
this is larger than the ``standard'' estimate
in Eq.\ (\ref{eq:Es}). The total radio energy per
SNR in \izw\ becomes
\begin{eqnarray}
{\cal E}^{\rm p}(0.1\mbox{--}100~{\rm GHz})=
1.5\mbox{--}6.0\times 10^{40}~{\rm J}\, .
\end{eqnarray}

\section{Synchrotron emission and magnetic fields \label{sec:mag}}

Through comparison with observations,
we have obtained the radio energy radiated by the
nonthermal synchrotron component. In this section, we
relate the observationally-derived values to the physical
quantities governing the synchrotron radiation.

\subsection{General estimate}

The energy spectrum of energetic
electrons radiating synchrotron radiation is
assumed to follow a power law as
\begin{eqnarray}
N(E)\,{\rm d}E=CE^{-p}\,{\rm d}E\, ,
\label{eq:energy}
\end{eqnarray}
where $N(E)\,{\rm d}E$ is the number density of the
energetic electrons with energies between $E$ and
$E+{\rm d}E$, and $C$ is the normalizing constant.
Under this energy spectrum, the energy of the energetic
electrons per unit volume,
$\varepsilon_{\rm e}$, is calculated as
\begin{eqnarray}
\varepsilon_{\rm e}=\int_{E_{\rm min}}^{E_{\rm max}}
EN(E)\,
{\rm d}E\, ,\label{eq:energy_tot}
\end{eqnarray}
where $E_{\rm max}$ and $E_{\rm min}$ are the
maximum and minimum energies of the energetic
electrons.
Combining equations (\ref{eq:energy}) and
(\ref{eq:energy_tot}), we can express the
normalizing constant as
\begin{eqnarray}
C & = & \left\{
\begin{array}{ll}
{\displaystyle
\frac{\varepsilon_{\rm e}(2-p)}
{E_{\rm max}^{2-p}-E_{\rm min}^{2-p}}
} & (p\neq 2) \\
{\displaystyle
\frac{\varepsilon_{\rm e}}{\ln (E_{\rm max}/E_{\rm min})}
} & (p=2)
\end{array}
\right. \, .
\end{eqnarray}

The emissivity of the synchrotron radiation,
$P(\nu )$, is written as
(Ginzburg \& Syrovatskii \cite{ginzburg65};
Rybicki \& Lightman \cite{rybicki79})
\begin{eqnarray}
P(\nu ) & = & 1.4\times 10^{-29}a(p)CB^{(p+1)/2}
\nonumber \\
& \times & \hspace{-2mm}\left(
\frac{\nu}{6.3\times 10^{18}~{\rm Hz}}
\right)^{-(p-1)/2}\,
{\rm W\,Hz^{-1}\, sr^{-1}},
\end{eqnarray}
where $a(p)$ is a coefficient which depends on
the exponent $p$ in the energy spectrum defined
in equation (\ref{eq:energy}), $C$ is the
normalizing constant in equation (\ref{eq:energy}),
and $B$ is the magnetic field strength in units of
gauss. 

The spectral index observed in individual SNRs is
roughly $\alpha\simeq -(p-1)/2\simeq -0.5$
(Clark \& Caswell \cite{clark76}); i.e., $p\simeq 2$.
Since the frequency of the power-law synchrotron
radiation ranges at least from MHz to GHz, we can
assume that $\ln (E_{\rm max}/E_{\rm min})\sim 10$.
In any case, because of the logarithmic dependence,
the results are not sensitive to the assumed
$E_{\rm max}/E_{\rm min}$.

The total nonthermal radio luminosity of a region
can be calculated by integrating $P(\nu )$ over all
the volume $V$ as
\begin{eqnarray}
L_{\rm nt}^{\rm theory}=\int_V4\pi P(\nu )\,{\rm d}V
\, ,
\end{eqnarray}
where $L_{\rm nt}^{\rm theory}$ is the theoretically
estimated nonthermal luminosity of the entire region.
Assuming homogeneity of the region and $p=2$, and
with $a(2)\simeq 0.103$
(Ginzburg \& Syrovatskii \cite{ginzburg65}) 
we obtain
\begin{eqnarray}
L_{\rm nt}^{\rm theory}(\nu ) & = & 2.0\times 10^{19}
\left(\frac{\ln (E_{\rm max}/E_{\rm min})}{10}
\right)^{-1} \nonumber \\
& \times & \left(
\frac{{\cal E}_{\rm e}}{10^{52}~{\rm erg}}\right)
\left(\frac{B}{10~\mu{\rm G}}\right)^{1.5} \nonumber \\
& \times & \left(\frac{\nu}{5~{\rm GHz}}\right)^{-0.5}~
{\rm W~Hz^{-1}}\, ,
\label{eq:lnt_theory}
\end{eqnarray}
where ${\cal E}_{\rm e}$ is the total energy in
the volume:
\begin{eqnarray}
{\cal E}_{\rm e}\equiv\int_V\varepsilon_{\rm e}\,
{\rm d}V\, .
\end{eqnarray}
We assume that a fraction 
$f_{\rm e}$ of the input
SN energy is converted to the energy of
accelerated energetic electrons:
\begin{eqnarray}
{\cal E}_{\rm e}\simeq f_{\rm e}E_{\rm SN}
\gamma\tau_{\rm e}\, ,
\end{eqnarray}
where $\tau_{\rm e}$ is the lifetime of energetic
electrons. Equating $L_{\rm nt,p}^0$
(Eq.\ \ref{eq:nt_para}) and $L_{\rm nt}^{\rm theory}$
(Eq.\ \ref{eq:lnt_theory}),
we obtain the theoretical
expression for $l_{\rm nt}\tau_{\rm nt}$ at
$\nu =5$ GHz as
\begin{eqnarray}
l_{\rm nt}\tau_{\rm nt} & = & 2.0\times 10^{19}
\left(\frac{\ln (E_{\rm max}/E_{\rm min})}{10}
\right)^{-1} \left(
\frac{f_{\rm e}E_{\rm SN}}{10^{47}\,{\rm erg}}
\right)\nonumber \\
& \times & \left(\frac{\tau_{\rm e}}{10^5\,{\rm yr}}
\right)
\left(\frac{B}{10\,\mu{\rm G}}\right)^{1.5}\,
{\rm W\, Hz^{-1}\, yr}\, .\label{eq:lt_theory}
\end{eqnarray}
It would be reasonable to assume that
$\tau_{\rm nt}=\tau_{\rm e}$. However, we retain
both quantities since the observational
constraint has been applied to
$l_{\rm nt}\tau_{\rm nt}$.

\subsection{Comparison with \sbs}
\label{subsec:lifetime}

In Sect.~\ref{subsec:sbs} for \sbs\ we derived the
nonthermal radio energy emitted at 5 GHz per SN as
$l_{\rm nt}\tau_{\rm nt}\simeq (6\mbox{--}22)\times 10^{22}$
W Hz$^{-1}$ yr. Comparing this value with the theoretical
expression given in Eq.\ (\ref{eq:lt_theory}), we obtain
$({f}_{\rm e}/10^{-4})(\tau_{\rm e}/10^5~{\rm yr})
(B/10~\mu{\rm G})^{1.5}\sim (3\mbox{--}11)\times 10^3$, where
we have assumed $E_{\rm SN}=10^{51}$ erg and
$\ln (E_{\rm max}/E_{\rm min})=10$.

Either $B\gg 10~\mu$G or $f_{\rm e}\gg 10^{-4}$ or both.
We first investigate the possibility that the
magnetic fields could be amplified by turbulence
generated from SNe as proposed by Balsara et al.\ (\cite{balsara04}).
We define the ratio of thermal energy to magnetic energy,
$\beta\equiv (3nk_{\rm B}T/2)/(B^2/8\pi )$, where
$n$ is the particle number density, $k_{\rm B}$ is
the Boltzmann constant, and $T$ is the gas
temperature. In \sbs, we can assume
$n\sim 10^3~{\rm cm}^{-3}$ and
$T\sim 2\times 10^4$ K (Sect.\ \ref{subsec:sbs}).
Then, the magnetic field strength becomes
$320\beta^{-1/2}~\mu$G.
The magnetic field could be amplified in the
star-forming region until the magnetic pressure
becomes comparable to the thermal pressure
(otherwise, the magnetic
fields would not be confined within the region).
Thus, $\beta >1$ would be a reasonable condition.
If we conservatively assume that $\beta\sim 10$,
we obtain the magnetic field strength in \sbs\ as
100 $\mu$G.
The production efficiency of energetic electrons
is not well known, but it could be assumed to be
of an order of 1\%  or more (e.g.,
Zirakashvili \& V\"{o}lk \cite{zirakashvili05}).
Here we conservatively assume
${f}_{\rm e}\sim 0.01$. With the quantities
estimated in this paragraph, we predict
$\tau_{\rm e}\sim (6-35)\times 10^4$ yr.

This timescale can be interpreted in two ways. One is
the radiative lifetime of a SNR. The adiabatic lifetime of a
SNR is estimated as $\tau_{\rm ad}\sim 10^3$ yr
if we assume $n_{\rm H}\sim 10^3$ cm$^{-3}$
(Eq.\ \ref{eq:adiabatic}). This is much shorter
than the above $\tau_{\rm e}$. However, the SNR
continues to expand after the cooling due to the
momentum conservation. If we adopt the expansion law
of Shull (\cite{shull80}), it finally reaches the
sound speed of the ionized medium and the shock
disappears after $\sim 10^5$ yr. Therefore, the
production of energetic electrons in
shocked gas may
continue on a timescale of $\sim 10^5$ yr, which is
roughly consistent with $\tau_{\rm e}$ estimated
above.
{From} the expansion law, we can derive
$\tau_{\rm e}\propto n^{-2/5}$.

The other interpretation of $\tau_{\rm e}$ is that
it reflects the lifetime of an energetic electron.
The energy loss of energetic electrons should be
considered at frequencies larger than $\nu_{\rm m}$
estimated as
\begin{eqnarray}
\nu_{\rm m}=\frac{3.1\times 10^{23}}{B_\perp^3t^2}~
{\rm Hz}\, ,\label{eq:nu_m}
\end{eqnarray}
where $B_\perp$ is the component of the magnetic
field perpendicular to the velocity of the electron,
and the time $t$ is measured from the production of
the energetic electrons
(Ginzburg \& Syrovatskii \cite{ginzburg65}).
We define $\tau_{\rm s}$ as the timescale on which
the energy loss of
electrons contributing to the radiation at a
frequency $\nu$ becomes significant.
Then, $\tau_{\rm s}$ can be estimated by inserting
$\nu_{\rm m}=\nu$ and $t=\tau_{\rm s}$ in
Eq.\ (\ref{eq:nu_m}) as
\begin{eqnarray}
\tau_{\rm s}\simeq 8.0\times 10^6\left(
\frac{B_\perp}{10~\mu{\rm G}}\right)^{-3/2}\left(
\frac{\nu}{5~{\rm GHz}}\right)^{-1/2}~{\rm yr}\, .
\label{eq:lifetime_syn}
\end{eqnarray}
If we adopt $B_\perp\sim 100~\mu$G for \sbs,
we obtain $\tau_{\rm s}\ga 2\times 10^5$ yr at
$\nu\la 10$ GHz. This is consistent with the
above $\tau_{\rm nt}$. In the following,
we adopt an approximation of $B_\perp\sim B$.

\subsection{Scaling of $l_{\rm nt}\tau_{\rm nt}$}
\label{subsec:scaling}

We investigate both cases for $\tau_{\rm nt}$. 
The first is
based on the radiative lifetime of a SNR, i.e.,
$\tau_{\rm nt}=\tau_{\rm e}\simeq 10^5
(n/10^3~{\rm cm}^{-3})^{-2/5}$ yr, and the
second on the lifetime of energetic
electrons, i.e., $\tau_{\rm nt}=\tau_{\rm s}$ given
in Eq.\ (\ref{eq:lifetime_syn}). By using
Eq.\ (\ref{eq:lt_theory}), these scaling relations
result in the following expression for
$l_{\rm nt}\tau_{\rm nt}$:
\begin{eqnarray}
l_{\rm nt}\tau_{\rm nt}\simeq\left\{
\begin{array}{l}
2.0\times 10^{21}\left(
{\displaystyle
\frac{f_{\rm e}E_{\rm SN}}{10^{49}~{\rm erg}}
}
\right)
\left({\displaystyle \frac{n}{10^3~{\rm cm}^{-3}}}
\right)^{-2/5} \\
~~~\times\left({\displaystyle \frac{B}{10~\mu{\rm G}}}
\right)^{1.5}\,{\rm W\, Hz^{-1}\, yr} \\
~~~\mbox{if $\tau_{\rm SN}=\tau_{\rm e}$,} \\
3.6\times 10^{23}\left({\displaystyle
\frac{f_{\rm e}E_{\rm SN}}{10^{49}~{\rm erg}}
}
\right)\,{\rm W\, Hz^{-1}\, yr} \\
~~~\mbox{if $\tau_{\rm SN}=\tau_{\rm s}$,}
\end{array}
\right.\label{eq:scaling_l_nt}
\end{eqnarray}
where we have assumed that
$\ln (E_{\rm max}/E_{\rm min})\simeq 10$.

It is generally difficult to observe the magnetic
field strength in BCDs. If we use $\beta$
(Sect.~\ref{subsec:lifetime}), we can express the
magnetic field strength in terms of the gas density as
$B\simeq 320~\mu{\rm G}\left(n/10^3~{\rm cm}^{-3}
\right)^{1/2}\beta^{-1/2}$, where we have used
$T=2\times 10^4$ K. A similar observational scaling
($B\propto n^{0.48}$) was observationally found by
Niklas \& Beck (\cite{niklas97}). 
Inserting this expression into
the first case in Eq.\ (\ref{eq:scaling_l_nt}),
we obtain a scaling relation as
$l_{\rm nt}\tau_{\rm nt}=8.1\times 10^{23}
(f_{\rm e}E_{\rm SN}/10^{49}~{\rm erg})
(n/10^3~{\rm cm}^{-3})^{0.35}\beta^{-0.75}$
W Hz$^{-1}$ yr. If we assume that
$f_{\rm e}E_{\rm SN}$ does not change among
galaxies and that $l_{\rm nt}\tau_{\rm nt}$ 
scales only with density, we obtain
$l_{\rm nt}\tau_{\rm nt}\simeq
(l_{\rm nt}\tau_{\rm nt})_{\rm SBS}
(n/10^3~{\rm cm}^{-3})^{0.35}$, where
$(l_{\rm nt}\tau_{\rm nt})_{\rm SBS}$ is the value
obtained for \sbs\ as reported in
Table \ref{tab:parameter}.

In the second case in Eq.\ (\ref{eq:scaling_l_nt}),
there is no dependence on magnetic fields and gas
density, if $f_{\rm e}E_{\rm SN}$ does not
depend on those quantities. Thus, we also consider
the case where $l_{\rm nt}\tau_{\rm nt}$ is independent
of $n$.

Finally we propose yet another scaling of the magnetic field
strength, assuming that the magnetic field is
amplified by SNe, as in the turbulence-driven amplification
mechanism proposed by Balsara et al.\ (\cite{balsara04}). 
Since the SN rate is proportional to the SFR, the energy density
of the magnetic fields could scale with the SFR
density ($\sim {\rm SFR}/r^3$).
Equation (\ref{eq:SFR_num}) implies that the
SFR density scales as $n^{3/2}$ (this scaling also
expresses a Schmidt law; Schmidt \cite{schmidt59}).
Thus, we obtain $B\propto n^{3/4}$.
With this scaling relation, and normalizing it to
\sbs, we obtain
$l_{\rm nt}\tau_{\rm nt}\simeq
(l_{\rm nt}\tau_{\rm nt})_{\rm SBS}
(n/10^3~{\rm cm}^{-3})^{0.73}$ for the first
case in Eq.\ (\ref{eq:scaling_l_nt}).
Since the energy density of energetic electrons is
proportional to the SN rate, this scaling indicates
that the magnetic energy density scales with
the energy density of energetic electrons.

In summary, the possible scaling of $l_{\rm nt}\tau_{\rm nt}$
is summarized by the following three cases:
\begin{eqnarray}
l_{\rm nt}\tau_{\rm nt}=\left\{
\begin{array}{ll}
(l_{\rm nt}\tau_{\rm nt})_{\rm SBS} &
\mbox{Scaling {\it a},} \\
(l_{\rm nt}\tau_{\rm nt})_{\rm SBS}
(n/10^3~{\rm cm}^{-3})^{0.35} &
\mbox{Scaling {\it b},} \\
(l_{\rm nt}\tau_{\rm nt})_{\rm SBS}
(n/10^3~{\rm cm}^{-3})^{0.73} &
\mbox{Scaling {\it c}.}
\end{array}
\right.\label{eq:scaling_summary}
\end{eqnarray}
Indeed, it
has recently been found that the $\Sigma$--$D$ (radio surface
brightness to diameter)
relation depends on the ambient density of interstellar medium
(Arbutina et al.\ \cite{arbutina04}; 
Arbutina \& Uro\v{s}evi\'{c} \cite{arbutina05}): SNRs with
higher ambient density tend to emit more radio energy.
Thus, the positive correlation between density and
$l_{\rm nt}\tau_{\rm nt}$ is supported.
We should note that thermal emission may contribute to the
radio emission from SNe especially in dense environments
(Uro\v{s}evi\'{c} \& Pannuti \cite{urosevic05}). Thus, our derived
values for $l_{\rm nt}\tau_{\rm nt}$ may include some
contribution from thermal emission. The decomposition
of thermal and nonthermal contributions is left for
future work.

\subsection{Comparison with \izw}

The typical number density of \izw\ is
$n\simeq 100~{\rm cm}^{-3}$ (Sect.~\ref{subsec:izw}).
Comparing Table~\ref{tab:parameter} (model for \sbs) with 
Tables~\ref{tab:param_izw12} and \ref{tab:param_izw15} (model
for \izw\ at 12 and 15\,Myr, respectively), we find that the
{\it best} model estimate for
$(l_{\rm nt}\tau_{\rm nt})_{\rm IZw}
/(l_{\rm nt}\tau_{\rm nt})_{\rm SBS}\simeq 0.13-0.33$, 
where the subscript ``IZw'' indicates the value for \izw. 
Scaling {\it a} would imply no density dependence, so that we would
expect 
$(l_{\rm nt}\tau_{\rm nt})_{\rm IZw}\simeq  (l_{\rm nt}\tau_{\rm nt})_{\rm SBS}$;
this is highly inconsistent with our models.
If, instead, we use Scaling {\it b} ($\propto\,n^{0.35}$),
we would derive
$(l_{\rm nt}\tau_{\rm nt})_{\rm IZw}
/(l_{\rm nt}\tau_{\rm nt})_{\rm SBS}\simeq 0.45$, 
still a bit larger than our model predictions.
Scaling {\it c} ($\propto\,n^{0.73}$) would give
$(l_{\rm nt}\tau_{\rm nt})_{\rm IZw}
/(l_{\rm nt}\tau_{\rm nt})_{\rm SBS}\simeq 0.19$, 
which is the most consistent with
our results, although Scaling {\it b} cannot be rejected with certainty.

Interestingly, the nonthermal
luminosity observed in discrete SNRs in ten external galaxies
appears to be positively correlated with density 
(Hunt \& Reynolds \cite{hunt06}),
in a way that is consistent with our results. 
If the time-luminosity integral varies as $n^{0.7}$, 
as in the seemingly most likely dependence, then
the observed correlation of $l_{nt} \propto n^{1.3}$ implies
that the radiative lifetime of a remnant should vary
as $n^{-0.6}$.
The slope of the $l_{nt} - n$ relation is rather uncertain
(Hunt \& Reynolds \cite{hunt06}), but cannot be less
than unity. 
With $l_{nt} \propto n$, the remnant lifetime would vary
as $n^{-0.3}$.
These values encompass the expansion lifetime dependence
$\tau_{\rm e}\propto n^{-2/5}$ and the adiabatic
one $\tau_{\rm e}\propto n^{-9/17}$.

\section{The time evolution of the radio continuum \label{sec:time}}

Cannon \& Skillman (\cite{cannon04}) propose the radio
spectral index as an age indicator.
The radio spectral index is defined by fitting the
radio spectrum with a functional form of
$\nu^{\alpha_{\rm f}}$. In their paper, the observational
spectral index is determined with the data at
$\nu =1.4$, 4.9, and 8.5 GHz.
They argue that $\alpha_{\rm f}>0$ with ages $\la 1$ Myr
because of the free-free absorption. Then the optically-thin thermal
spectral index ($\alpha_{\rm f}=-0.1$) appears
within a typical lifetime of \hii\ regions
($\sim 10$ Myr) and finally the spectrum steepens because
of the increasing contribution from a synchrotron component.

\begin{figure*}
\centering
\includegraphics[width=8cm]{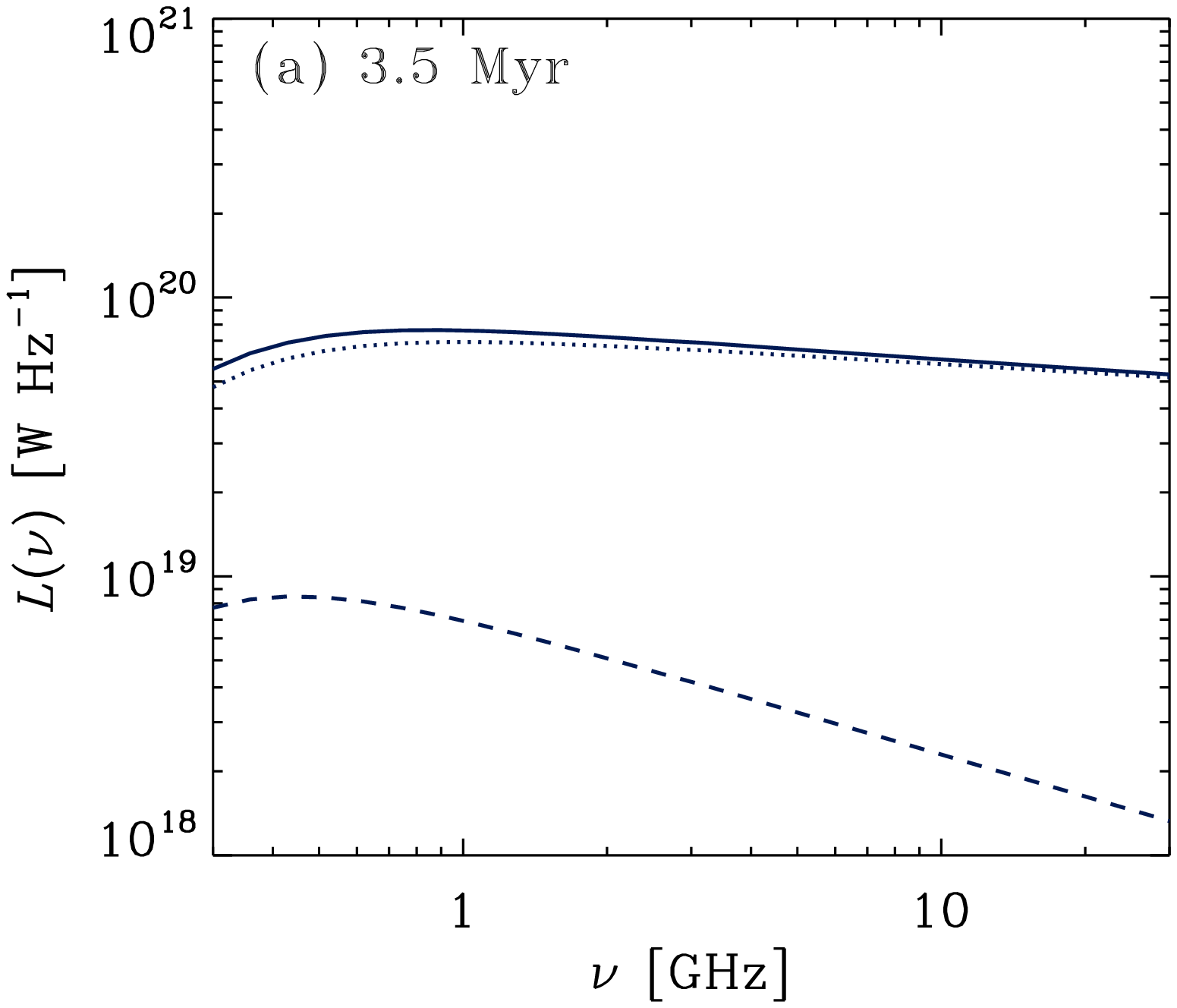}
\includegraphics[width=8cm]{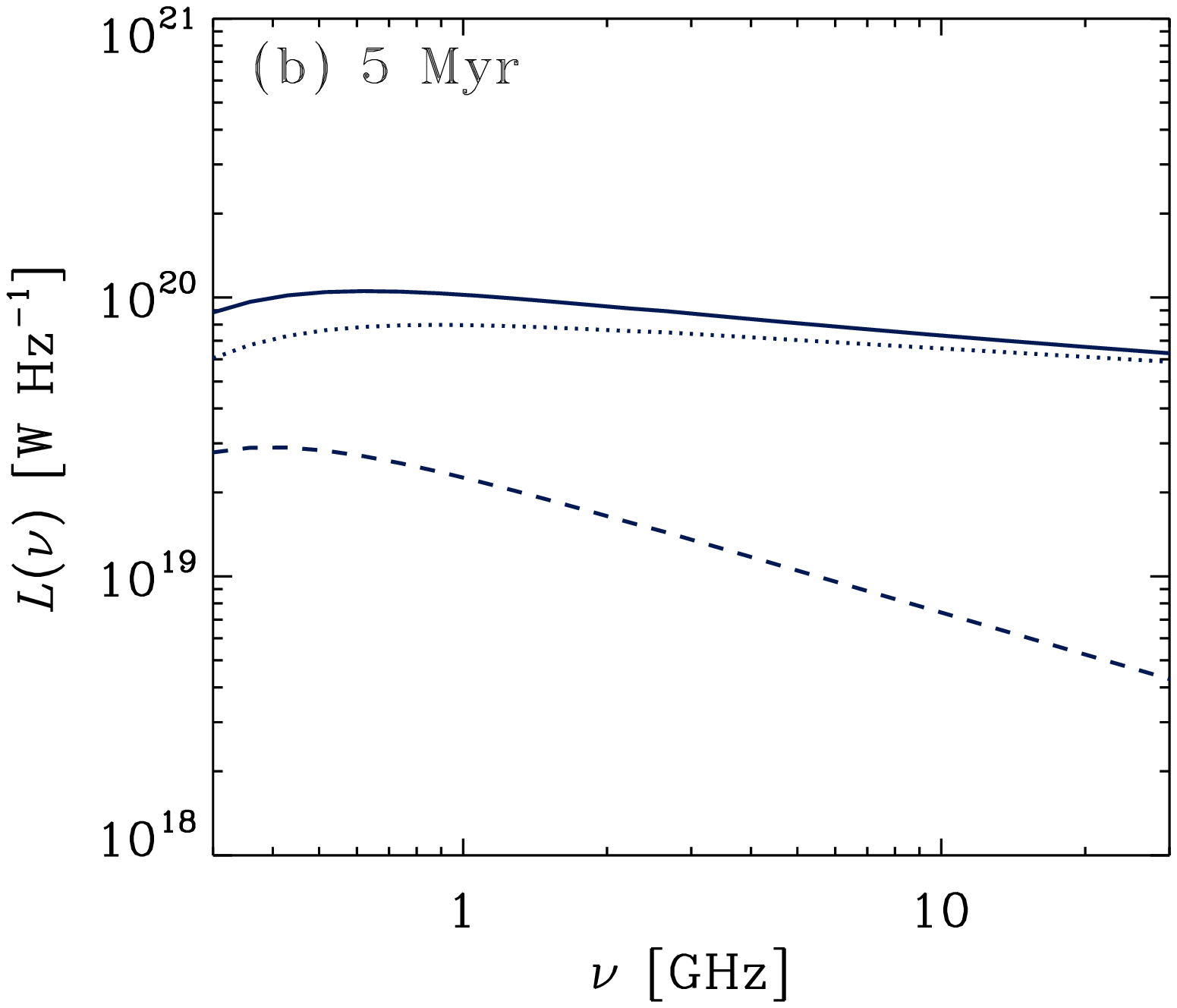}\\
\includegraphics[width=8cm]{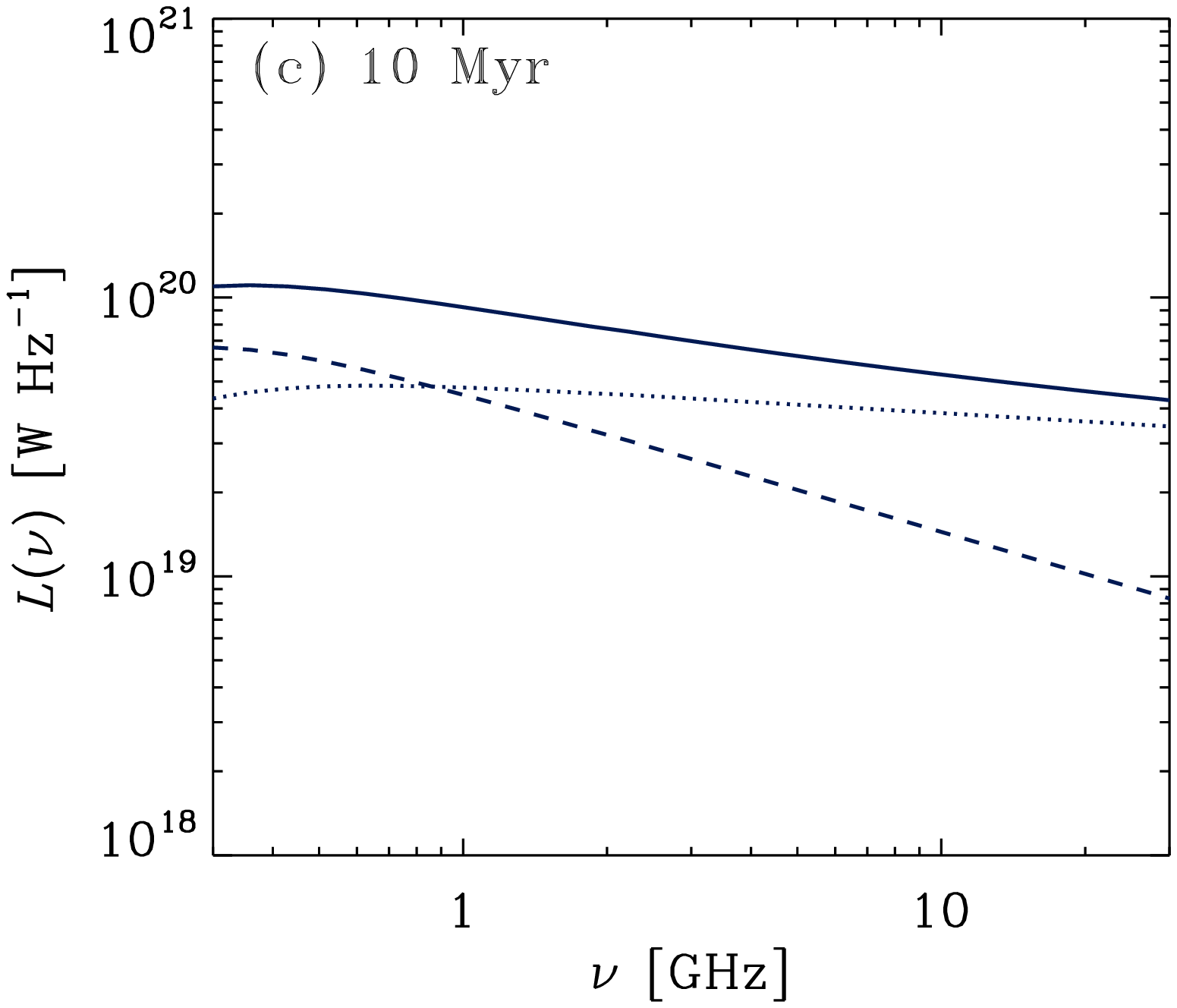}
\includegraphics[width=8cm]{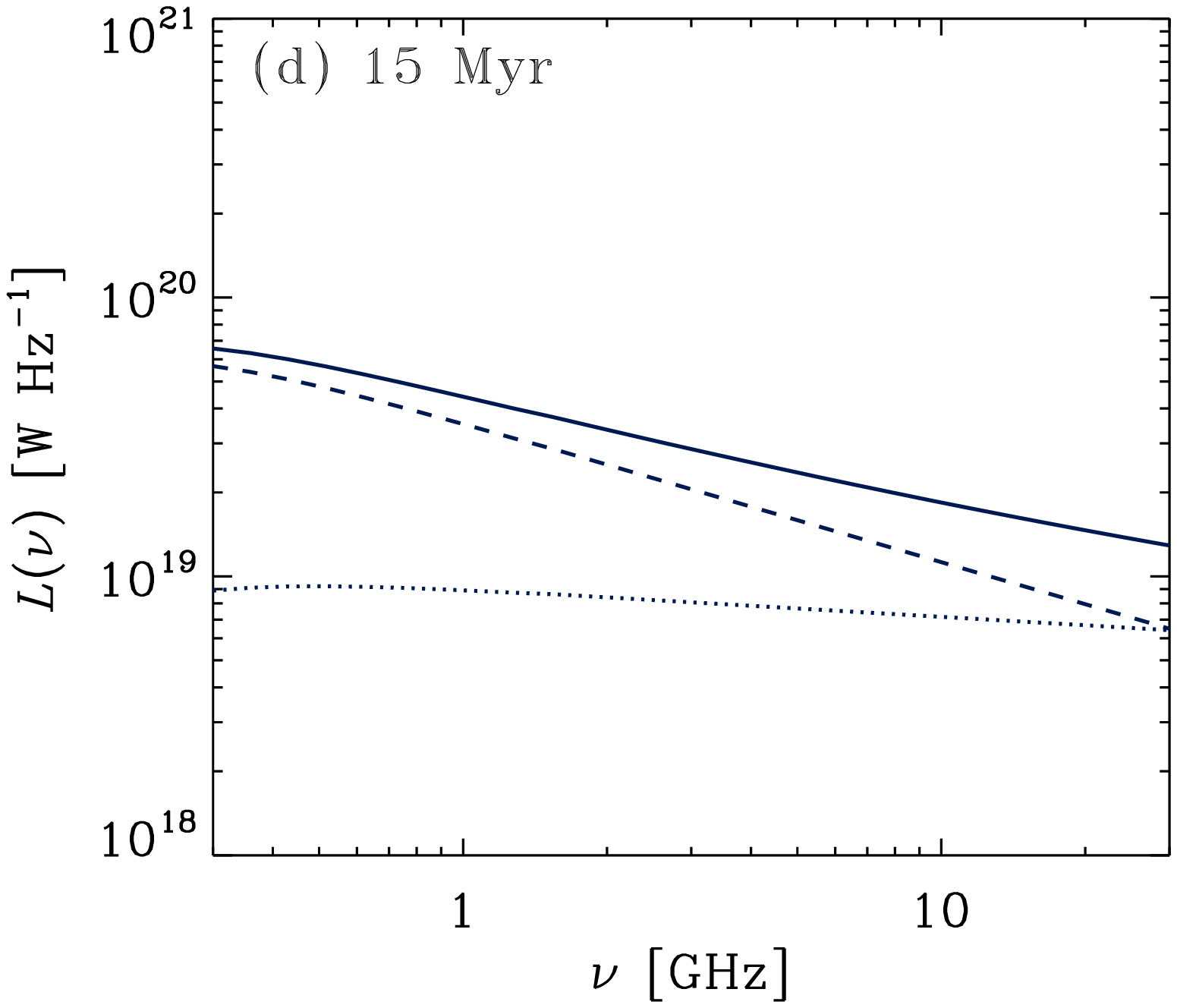}
\caption{Same as Fig.\ \ref{fig:ev_active}
but for the ``passive'' mode, with the \izw\ model as a
representative case.}
\label{fig:ev_passive}
\end{figure*}

However, we have shown in Sects.\ \ref{subsec:sbs} and
\ref{subsec:izw} that the gas density is also important for
the shape of
the radio spectrum. The dense and compact
star-forming region in \sbs\ shows a strong burst of
star-formation with strong free-free absorption. We have
dubbed this characteristic ``active''. On the other hand,
the diffuse star-forming region in \izw\ has more mild
star-formation with much less absorption. We have called this mode
``passive''. Since the radio free-free absorption
affects the radio spectral index, we can distinguish these
two classes of star formation, ``active'' and ``passive'',
through the radio spectral index. This also means that the
age is not the only factor that determines the spectral
index; the gas density is also important in its determination. 

Here we examine the time evolution of radio spectra, adopting
the same parameters as used for \sbs\ and \izw, considering them
as representative cases for ``active'' and ``passive'',
respectively. Model G is adopted here, but the qualitative
behavior of the radio spectral index is not much affected
by the specific details of the adopted models. 
We use $l_{\rm nt}\tau_{\rm nt}$ estimated for Model G
in Tables \ref{tab:parameter} and
\ref{tab:param_izw12} for the active and passive cases,
respectively. In Figs.\ \ref{fig:ev_active} and
\ref{fig:ev_passive}, we present the radio SEDs calculated
by the models for \sbs\ and \izw, respectively.
The snapshots at $t=3.5$ Myr (soon after the first explosion
of SNe at $\simeq 3$ Myr), 5 Myr, 10 Myr, and 15 Myr are
shown. We observe that the free-free absorption around
$\nu =1$ GHz is prominent in the active case even at
$t\ga 5$ Myr.
On the contrary, there is little absorption even at
$t\sim 3.5$ Myr in the passive case. This indicates that
the absorption feature is not a simple indicator of
age as suggested by Cannon \& Skillman (\cite{cannon04}).
The density of the star-forming region is also important for
the spectral index around $\nu =1$ GHz.

By comparing Figs.\ \ref{fig:ev_active} and
\ref{fig:ev_passive}, we also see that the nonthermal
component becomes comparable to the thermal luminosity
earlier in the active case than in the passive case.
This mainly comes from the difference in time-integrated
radio luminosity of a SN ($l_{\rm nt}\tau_{\rm nt}$).
Thus, we suggest that a nonthermal-dominated radio SED
is observed more often in active BCDs than in passive
BCDs. This is true if $l_{\rm nt}\tau_{\rm nt}$
correlates with gas density.
Observations compiled by Hunt \& Reynolds (\cite{hunt06}) 
support this view.

\begin{figure}
\includegraphics[width=8cm]{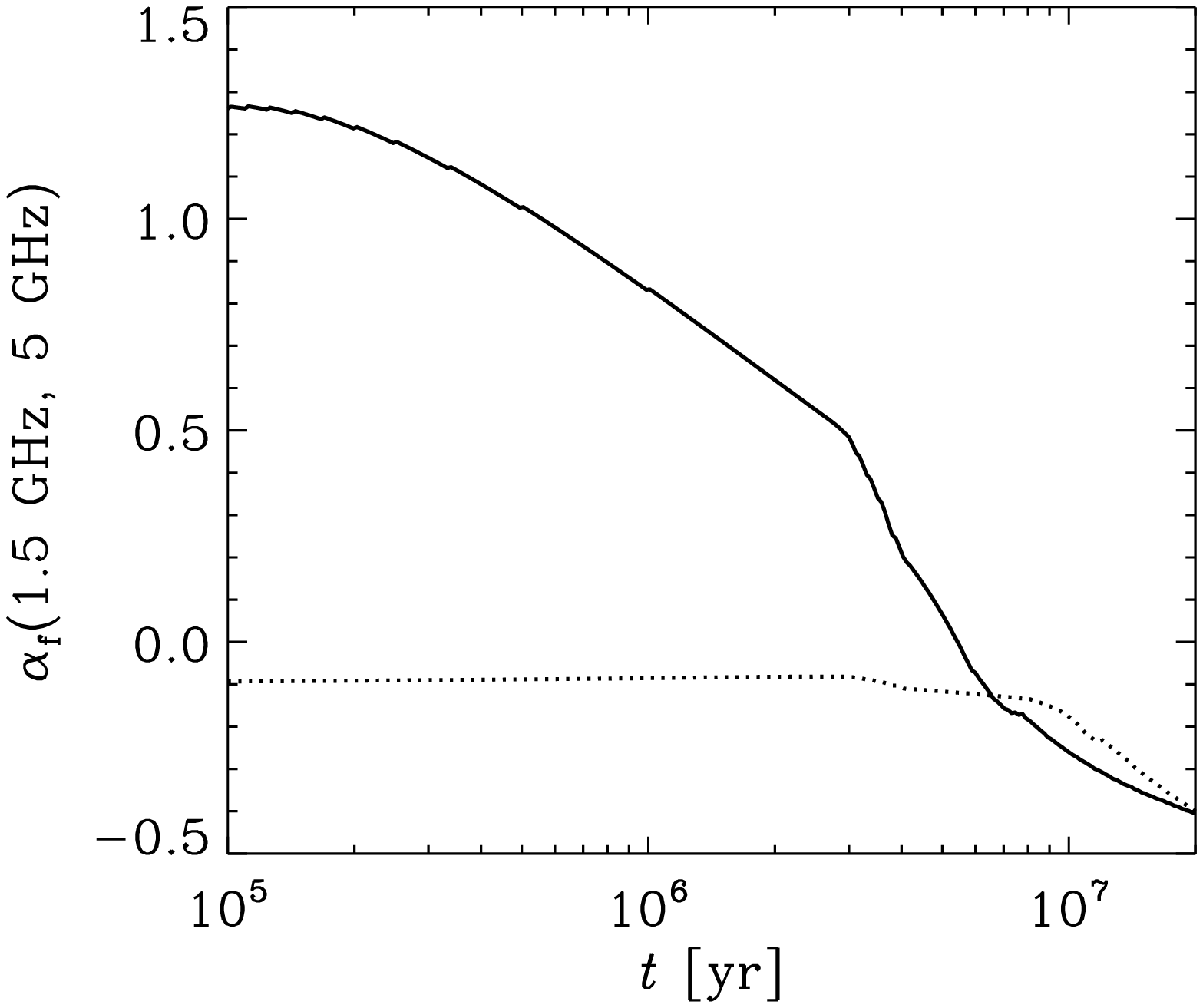}
\caption{Time evolution of the radio spectral
index defined at 1.5 GHz and 5 GHz
[$\alpha (1.5~{\rm GHz},\, 5~{\rm GHz})$].
The solid and dotted lines show the results
of the ``active'' and ``passive'' models
(same as Figs.\ \ref{fig:ev_active} and
\ref{fig:ev_passive}, respectively).}
\label{fig:index}
\end{figure}

Finally we show the time evolution of radio spectral index
$\alpha_{\rm f}$. The spectral index defined at
$\nu =\nu_1$ and $\nu_2$ is calculated by the following
equation:
\begin{eqnarray}
\alpha_{\rm f}(\nu_1,\,\nu_2)=
\frac{\log [L(\nu_1)/L(\nu_2)]}{\log (\nu_1/\nu_2)}\, .
\end{eqnarray}
We put $\nu_1=1.5$ GHz and $\nu_2=5$ GHz to be consistent
with the frequency range often used to determined the
spectral index.Figure \ref{fig:index} shows the
results. The solid and dotted lines represent the results
of the ``active'' and ``passive'' models whose spectral
shapes are shown in Figs.\ \ref{fig:ev_active} and
\ref{fig:ev_passive}, respectively.
The spectral index depends on age, and it approaches
the nonthermal value ($-0.4$) as the age increases.
This confirms the trend suggested by
Cannon \& Skillman (\cite{cannon04}). 
More importantly though, the spectral index is not a simple
function of age but rather depends strongly on the
density of the star-forming region (i.e.,
``active'' and ``passive'' modes). Indeed the spectral
index is positive even at $t=\mbox{a few Myr}$
in the active mode, while
it is always negative in the passive mode.

Another important consideration is that the radio spectrum
can be flat (i.e., $\alpha_{\rm f}\sim 0$)
even when the contribution from the nonthermal
component is significant. Indeed, at $t=5$ Myr, the
nonthermal and thermal
components are comparable in the
active mode (Fig.\ \ref{fig:ev_active}), and the
spectral slope at high frequency indicates the presence
of the nonthermal contribution. However, if we define
the spectral index by using $\nu =1.5$ GHz and 5 GHz,
the spectral index becomes $\sim 0$. This spectral
index could be misinterpreted as thermal. Thus,
in order to avoid such a misinterpretation, it is
important to derive a spectral index at $\nu\ga 5$ GHz.

\subsection{Implications for primeval galaxies}

High-$z$ primeval galaxies tend to have a high
density; they are formed and evolve in deep gravitational 
potentials (e.g., Padmanabhan \cite{padmanabhan93}).
Therefore, the gas in such primeval objects may
have high pressure, and the star-forming regions
may be expected to mimic the active mode
(Hirashita \& Hunt \cite{hirashita04}).
Indeed, the number density of gas in high-$z$ galaxies
is estimated as $\ga 10^3$ cm$^{-3}$
(Norman \& Spaans \cite{norman97}), and is similar
to that of \sbs, typical of the active class.
Therefore, the
nonthermal component could dominate
on short timescales after the onset of star formation
while the radio emission is optically thick at
$\nu\la 1$ GHz. The dense environment could also aid
amplification of magnetic fields as proposed by the
scaling relations in Sect.\ \ref{subsec:scaling}.

This paper also has implications for the production
of high-energy electrons. It would be reasonable to
consider that the production occurs on an expansion
timescale SNRs, since the shocks associated with SNRs
can be responsible for the acceleration of electrons
(e.g., Axford \cite{axford81}). The lifetime of
nonthermal radiation per SN is determined by the
minimum of two timescales: the lifetime of the shock and
the synchrotron loss timescale.

\section{Conclusions}\label{sec:conclusion}

In order to investigate the time evolution of the
radio SEDs of metal-poor star-forming galaxies, we have
constructed a radio SED model by
treating the thermal and nonthermal components
consistently with the star formation history.
In particular, the
duration and luminosity of the
nonthermal radio emission from supernova remnants
(SNRs) has been constrained by using the observational radio SEDs
 of \sbs\ and \izw, both of which are reasonable proxies for
young galaxies
in the nearby universe. In \sbs, the typical radio energy emitted
per SNR over its lifetime is estimated to be
${\cal E}(0.1-100~{\rm GHz})\sim 8\mbox{--}32
\times 10^{47}~{\rm erg}$. In
\izw, another representative of metal-poor star-forming
galaxies, we find that
${\cal E}(0.1-100~{\rm GHz})\sim 2\mbox{--}6\times 10^{47}$ erg. 
Both estimates are significantly larger than
previous estimates based on the $\Sigma-D$ relation.
These values of the two ``template'' galaxies can be
simultaneously explained by a simple density scaling
relation of $B\propto n^{0.35-0.73}$, indicating that
the magnetic pressure scales with the gas pressure,
or that the magnetic fields are amplified
by SNe, or both.

We have also predicted the time dependence of the radio
spectral index and of the radio SED itself, for both
the active (\sbs) and passive (\izw) cases.
These models enable us to roughly age date and classify
radio spectra of star-forming galaxies into active/passive classes.
Since the radio emission around $\nu\la 1$ GHz can be
affected by the free-free absorption especially in the
active class, the spectral index defined at
$\nu\ga 5$ GHz should be used to estimate the contribution
from the nonthermal component. On the other hand,
the free-free absorption feature around $\nu\sim 1$ GHz
could be used to select the active class.

\begin{acknowledgements}

We thank the anonymous referee for useful comments which improved
this paper considerably. We are grateful to T. Yoshida,
D. Uro\v{s}evi\'{c}, Y. Kato,
T. Kamae, and H. Kamaya for stimulating discussions on supernovae,
high energy phenomena,
and magnetic fields. HH has been supported by the University of Tsukuba
Research Initiative and by Grants-in-Aid for Scientific
Research of the Ministry of Education, Culture, Sports,
Science and Technology (Nos.\ 18026002 and 18740097).
This research has made use of the NASA/IPAC
Extragalactic Database (NED), which is operated by the Jet Propulsion
Laboratory, California Institute of Technology, under contract
with the National Aeronautics and Space Administration (NASA).
We have relied upon NASA's Astrophysics
Data System Abstract Service (ADS).

\end{acknowledgements}

\appendix

\section{Analysis of expanding H {\sc ii} regions}
\label{app:analytic}

We derive some approximate analytic solutions for
expanding ionized regions based on
Eqs.\ (\ref{eq:expand1}) and (\ref{eq:expand2}).
By combining these two equations,
we obtain
\begin{eqnarray}
\frac{{\rm d}r_{\rm i}}{{\rm d}t}=C_{\rm II}\left(
\frac{r_{\rm i0}}{r_{\rm i}}\right)^{3/4}\, .
\label{eq:dri}
\end{eqnarray}
Since we estimate $r_{\rm i0}$ by the Str\"{o}mgren
radius under the current
stellar luminosity and the gas density of
the neutral medium outside the ionized region,
we obtain
\begin{eqnarray}
r_{\rm i0}={\cal A}\,
\frac{\dot{N}_{\rm ion}(t)^{1/3}}{n_{\rm H0}^{2/3}}\, ,
\label{eq:ri0}
\end{eqnarray}
where
\begin{eqnarray}
{\cal A}\equiv\left(
\frac{3}{4\pi(1.08)\alpha^{(2)}}\right)^{1/3}\, .
\label{eq:calC}
\end{eqnarray}
The factor 1.08 comes from the correction for ionized
helium (Eq.\ \ref{eq:corr_he}). By combining
Eqs.\ (\ref{eq:dri})--(\ref{eq:calC}),
we obtain
\begin{eqnarray}
\frac{{\rm d}r_{\rm i}}{{\rm d}t}=C_{\rm II}
{\cal A}^{3/4}
n_{\rm H0}^{-1/2}[\dot{N}_{\rm ion}(t)]^{1/4}
r_{\rm i}^{-3/4}\, .
\end{eqnarray}
Solving this equation, we obtain
\begin{eqnarray}
\frac{4}{7}r_{\rm i}^{7/4}=C_{\rm II}{\cal A}^{3/4}
n_{\rm H0}^{-1/2}\int_0^t N_{\rm ion}(t')^{1/4}
{\rm d}t'\, ,
\label{eq:ri_analytic}
\end{eqnarray}
where we have assumed that $r_{\rm i}(t=0)=0$.
If we assume that the SFR is
constant and that the death of stars is negligible,
$\dot{N}_{\rm ion}$ is almost proportional to
$t$, and we can predict that
$r_{\rm i}\propto t^{5/7}$.
This relation, combined with Eqs.\ (\ref{eq:expand1})
($\rho_{\rm II}\propto ({\rm d}r_{\rm i}/{\rm d}t)^2$),
indicates
that $\rho_{\rm II}\propto t^{-4/7}$. The emission
measure evolves as
${\rm EM}\propto\rho_{\rm II}^2r_{\rm i}\propto t^{-3/7}$.
In spite of this rough derivation, those relations explain
the behaviors depicted in Fig.\ \ref{fig:sbs_basic}.
In the burst SFR (Fig.\ \ref{fig:sbs_basic}),
the emission measure is particularly deviates from the
predicted power-law evolution. This is because the
first increase of $r_{\rm i}$ is due to increase
of the Str\"{o}mgren radius itself, not due to
dynamical expansion.


\end{document}